%% file: 2SIS.tex
\documentclass[usenatbib]{mn2e}

\usepackage{xspace}
\usepackage{graphicx}
\usepackage{amsmath}
\usepackage{caption}
\usepackage{xcolor}
\newcommand{\david}{}
\input{macros.tex}
\input{addresses.tex}

\date{Submitted to MNRAS}

\title[Compound lensing: Einstein Zig-Zags and high multiplicity lensed images]
{Compound lensing: Einstein Zig-Zags and high multiplicity lensed images}
    
\author[Collett \etal]{%
  Thomas~E.~Collett\thanks{\collettemail}, David~J.~Bacon
  \\\icg
}

\begin{document}

\pagerange{\pageref{firstpage}--\pageref{lastpage}}\pubyear{2015}

\maketitle 

\label{firstpage}

\begin{abstract}

Compound strong gravitational lensing is a rare phenomenon, but a handful  of such lensed systems are likely to be discovered in forthcoming surveys. In this work, we use a double SIS lens model to analytically understand how the properties of the system impact image multiplicity for the final source. We find that up to six images of a background source can form, but only if the second lens is multiply imaged by the first and the Einstein radius of the second lens is comparable to, but does not exceed that of the first. We then build a model of compound lensing masses in the Universe, using SIE lenses, and assess how the optical depth for multiple imaging by a galaxy-galaxy compound lens varies with source redshift. For a source redshift of 4, we find optical depths of $6 \times 10^{-6}$ for multiple imaging and $5 \times 10^{-8}$ for multiplicity of 6 or greater. We find that extreme magnifications are possible, with magnifications of 100 or more for  $6 \times 10^{-9}$  of $z=10$ sources with 0.1 kpc radii. We show some of the image configurations that can be generated by compound lenses, and demonstrate that they are qualitatively different to those generated by single-plane lenses; dedicated compound lens finders will be necessary if these systems are to be discovered in forthcoming surveys.

\end{abstract}


\setcounter{footnote}{1}


\section{Introduction}
\label{sec:intro}

Strong gravitational lensing is a rare phenomenon, but forthcoming surveys are forecast to discover hundreds of thousands of strong lenses in the next decade \citep{collett2015a}. This huge strong lens population will include a subset of -- extremely rare -- exotic strong lenses, where higher order lensing catastrophes create six or more images of a single background source. Exotic lenses are primarily interesting for three reasons: they are often powerful cosmic telescopes \citepeg{orban,wong} with extreme source magnifications near the catastrophes; the image configuration is extremely sensitive to the dark matter distribution in the lens(es) \citepeg{orban,sonnenfeld2012}, and if the lenses are at different redshifts, or the sources are time-variable they are powerful cosmological probes \citepeg{collett2012,collett2014}.

The simple picture of lensing by an elliptically symmetrical mass distribution with monotonically decreasing density gives rise to at most five images of a background point source \citep{evans+witt}. However when multiple galaxies play a significant role in the lensing, the formation of additional images \citepeg{keetonmaowitt} is possible. Recently several exotic lenses have been discovered; at cluster scales systems with more than five images are not uncommon. For example SDSS J2222+2745 is a cluster lensing a quasar into six images \citep{dahle2013}, the cluster-lensed supernova {\it ``Refsdal''} is predicted to be sextuply imaged \citep{snrefsdal}, and Abell 1703 contains an ultra-rare hyperbolic umbilic catastrophe \citep{orban}. For less massive systems, sextuples are harder to create but can occur for binary lenses \citepeg{shin+evans} or for individual galaxies with multipolar structures \citep{evans+witt}, such as galaxies with significant bulge and disc components \citep{orban}. 

Theoretical work on understanding lensing exotica has focused on systems with a single lens plane; however it is also possible for exotica to occur in compound lenses. \citet{kochanek+apostolakis} first considered lensing by two isothermal spheres on different planes; finding seven image configurations for certain lens and source parameters, but their investigation was limited by the computational power then available. \citet{werner} considered the same model and showed that it is possible to create multiple Einstein rings of a background source, if all three objects lie on a single optical axis, and the Einstein radius of the two lenses are comparable. \citet{moeller} also considered compound lensing by galaxies, finding that roughly 1 in 20 multiply imaged sources will be a compound lens with merging caustics.

 A handful of galaxy-galaxy-galaxy lenses are now known \citep{slacsVI, belokurov}. Early work on the double source plane lens SDSSJ0946+1006 by \citet{slacsVI} and \citet{sonnenfeld2012} preferred non-zero lensing effect by the first source acting on the second. Recently a full reconstruction of the lensed arcs by \citet{collett2014} measured the Einstein radius of the first source to be $0.15 \pm 0.02$ arcseconds, even without fixing the cosmological parameters; this demonstrates that there is indeed a compound lensing effect in this system.

The masses and positions in SDSSJ0946+1006 give rise to nothing more exotic than an Einstein ring from each source, but it is likely that the hundreds of thousands of strong lenses discoverable in future surveys will contain exotic compound lenses. In this work we aim to understand the configurations, magnifications and abundances of exotic compound lenses that are likely to be discovered in the near future.

In this work we are primarily concerned with answering three questions:
\begin{itemize}
  \item How often does compound lensing generate high multiplicity systems?
  \item How often does compound lensing generate extreme magnifications?
  \item What do high multiplicity/magnification compound lenses look like?
\end{itemize}

{\tom The paper is organised as follows. In Section \ref{sec:theory}, we review the general theory of compound lensing, which we apply to the case of two singular isothermal sphere lenses in Section \ref{sec:2SIS}. We provide analytical results for the case of the observer, lenses and source lying on a plane in section \ref{sec:axis}, and show numerical results for the general system in section \ref{sec:offaxis}. In Section \ref{sec:2SIE} we investigate the effect of lens ellipticities on the system, and calculate optical depths for high multiplicity and high magnification compound lensing. We discuss our results, and answer the primary questions above in Section \ref{sec:conclude}.}

\section{Compound lensing}
\label{sec:theory}

In this section we briefly review the theory of compound lensing, following \citet{sef}.

For a single-source-plane lens, the lens equation can be written as
\begin{equation}
\label{eq:SSPL}
\mathbf{\hat{x}}_2 = D_2/D_1 \mathbf{\hat{x}}_1 - D_{12}\vect{\hat{\alpha}}_1(\mathbf{\hat{x}}_1),
\end{equation}
where $\mathbf{\hat{x}_1}$ is the physical position vector in the image plane, $\mathbf{\hat{x}_2}$ is the unlensed source position vector, and ${\vect{\hat{\alpha}_1}(\mathbf{\hat{x}_1})}$ is the physical deflection a ray undergoes as it passes through $\mathbf{\hat{x}_1}$. $D_{\mu \nu}$ are angular diameter distances between planes $\mu$ and $\nu$. Once a light ray has been traced back to the second plane, we can trace it back to the third, remembering that the second deflection is a function of where it passes through the second plane. Recursively repeating this allows us to write down the multiple-plane lens equation
\begin{equation}
\label{eq:SSPL}
\mathbf{\hat{x}}_\nu = D_\nu/D_1 \mathbf{\hat{x}}_1 - \sum_{\mu=1}^{\nu-1}D_{\mu\nu}\vect{\hat{\alpha}}_\mu(\mathbf{\hat{x}}_\mu).
\end{equation}
The multiple-plane lens equation is more elegantly written in angular quantities by re-scaling the physical deflections to their angular effect on a ray from the final source plane,
\begin{equation}
\vect{\alpha}_\mu = \frac{D_{\mu s}}{D_{s}}\hat{\vect{\alpha}}_\mu.
\end{equation}
This gives
\be
\label{eq:multilensequation}
\vect{x}_\nu=\vect{x}_1 - \sum_{\mu=1}^{\nu-1}\beta_{\mu\nu}\vect{\alpha}_{\mu}(\vect{x}_\mu),
\ee
where $\vect{x}_\nu$ is the angular position on plane $\nu$. $\beta_{\mu\nu}$ is defined as
\begin{equation}
  \beta_{\mu\nu} \equiv \frac{D_{\mu\nu} D_{s}}{D_{\nu} D_{\mu s}}.
\end{equation}
Since in this paper we are only interested in the case of a three plane system (see Figure \ref{fig:setup}), we will use the shorthand 
\begin{equation}
  \beta \equiv \frac{D_{12} D_{3}}{D_{2} D_{23}}.
\end{equation}
In the case of a three plane system, the lens equation for photons originating on the second plane is
\label{eq:DSPLs1}
\begin{equation}
\mathbf{y} = \mathbf{x} - \beta \vect{\alpha_{\rm {1}}}(\mathbf{x}),
\end{equation}
where $\mathbf{y}$ is the unlensed position of the source on plane 2.
For photons originating on the third plane,
\begin{equation}
\mathbf{z} = \mathbf{x} - \vect{\alpha_{\rm {1}}}(\mathbf{x}) - \vect{\alpha_{\rm {2}}}(\mathbf{x}- \beta \vect{\alpha_{\rm {1}}}(\mathbf{x})),
\label{eq:DSPLs2}
\end{equation}
where $\mathbf{z}$ is the unlensed position of the source on plane 3.

Throughout this work we will refer to angular positions on the image plane, $\mathbf{x_1}$, as
\be
\mathbf{x_1}=\vecarr{x_i \\ x_j}=\vecarr{r \cos(\theta) \\ r \sin(\theta)}
\ee
We will work in units scaled to $\alpha_1 = 1$ unless otherwise specified.

\section{Compound lensing by two isothermal spheres}
\label{sec:2SIS}

\begin{figure}
  \centering
    \includegraphics[width=\columnwidth,clip=True]{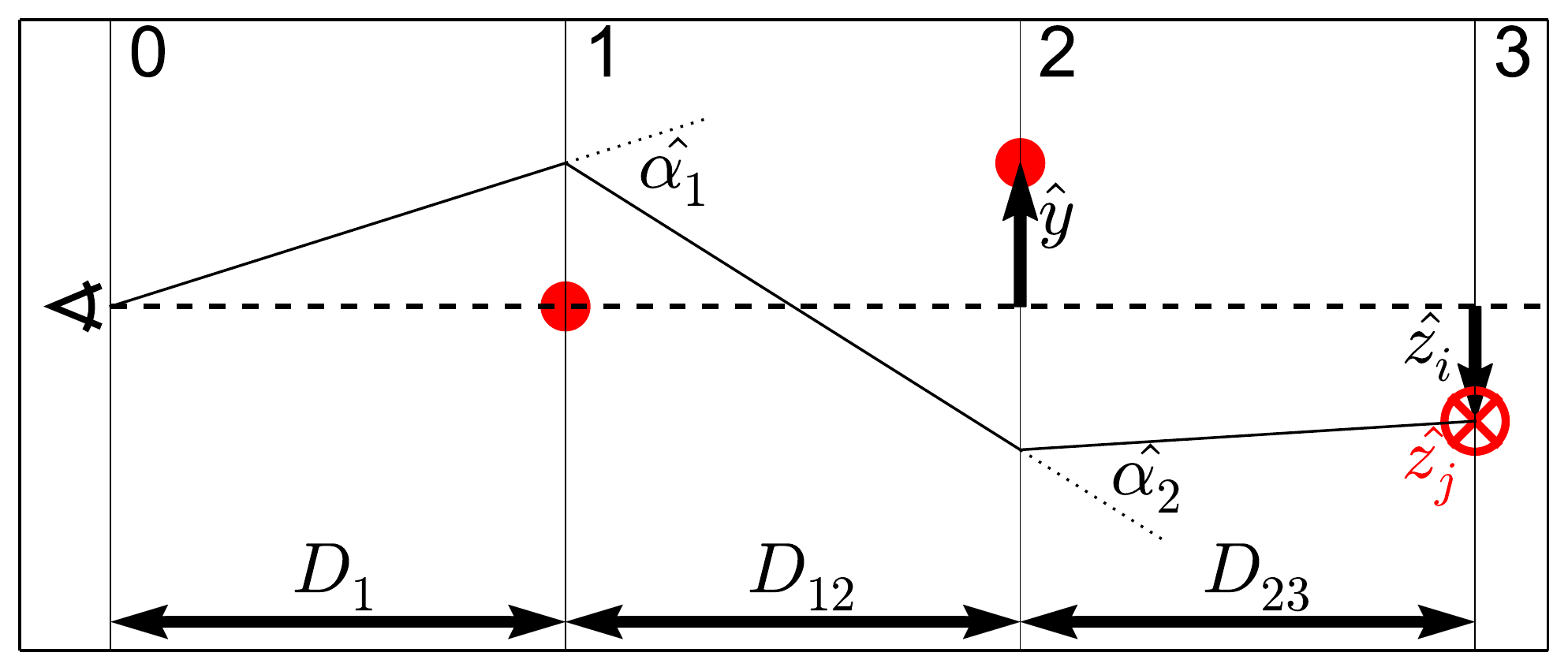}
    \caption { The {\david configuration considered for}  a two component lens. {\david The observer O is set at the far left of the figure.} The primary lens lies on the optical axis, with angular coordinates $(0,0)$; the location of the second lens defines the second axis and lies at $(y,0)$. The source lies at  $(z_i, z_j)$ with the $z_j$ coordinate in the direction perpendicular to the plane of the page. Quantities shown on the diagram are physical (hence hatted).}
    \label{fig:setup}
\end{figure}

For simplicity we focus primarily on the case where both lenses are singular isothermal spheres (SISs). For this model we can analytically derive some key results. A schematic of our system is shown in Figure \ref{fig:setup}.
Elliptical galaxies dominate the strong lensing cross-section, and have been shown to be approximately isothermal in their total density profile \citep{slacsX}, so the analytic results for the double SIS model can give us intuition for more general compound lensing.

The deflection angles of an SIS have the analytically convenient property that their magnitude is constant, and the direction always points at the lensing centre:
\be
\vect{\alpha}(\mathbf{r})=\alpha \hat{\vect{r}}.
\ee
Symmetry allows us to place the first lens on the optical axis, and the second lens at 
\be
\mathbf{y}=\vecarr{y \\ 0},
\ee
insisting that $y>0$. The light associated with lens 2 is thus observed at $\mathbf{x_1}=(x_i, 0)$
\be
y= x_i - \beta \alpha_1 x_i/|x_i|. 
\ee
For $x>0$, this gives a solution with an image at 
\be
x_i = y+ \beta \alpha_1,
\ee
which always exists since $y$, $\beta$ and $\alpha_1$ are all positive. A second solution exists for $x_i<0$ at
\be
x_i = y - \beta \alpha_1,
\ee
if and only if 
\be
\label{eq:SIScon1}
y< \beta \alpha_1.
\ee 
Equation \ref{eq:SIScon1} is thus the condition for multiple imaging of the { lens residing} on plane 2.

The final source is at $\mathbf{z}=(z_i, z_j)$.
By symmetry we can insist that $z_j>0$, but $z_i$ can take any value. The deflection caused by the second SIS is given by
\be
\vect{\alpha}_2(\mathbf{x}_2)=\alpha_2 \widehat{\left(\mathbf{x}_2-\mathbf{y}\right)}
\ee
Since $\mathbf{x}_2=\mathbf{x}_1 - \beta \alpha_1 \hat{\mathbf{x}}_1$, substituting into Equation \ref{eq:DSPLs2} quickly becomes unwieldy. It is much more convenient to write $\mathbf{x}_1$ in polar coordinates, centred on the first lens,
\be
\mathbf{x}_1=\vecarr{r \cos{\theta}\\ r \sin{\theta}},
\ee
giving
\be
\mathbf{x}_2=\vecarr{\cos{\theta}\\\sin{\theta}}(r-\beta \alpha_1)
\ee
and
\be
\widehat{\mathbf{x}_2-\mathbf{y}}= 
\frac{r-\beta \alpha_1}{D} \vecarr{\cos{\theta}\\\sin{\theta}}-\frac{1}{D}\vecarr{y\\0}
\ee
with
\be
\label{eq:D}
D=\left((r-\beta \alpha_1)^2+y^2-2y(r-\beta \alpha_1)\cos{\theta} \right)^{1/2}.
\ee
Substituting into Equation \ref{eq:DSPLs2} gives
\be
\label{eq:dsplequation2}
\vecarr{z_i \\z_j}=\left(r-\alpha_1-\frac{\alpha_2(r-\beta \alpha_1)}{D}\right)\vecarr{\cos{\theta}\\\sin{\theta}}-\frac{\alpha_2}{D}\vecarr{y\\0}.
\ee
Equation \ref{eq:dsplequation2} now allows us to solve for the location and multiplicity of images for any source position.

\subsection{Analytic solutions for the case $z_j=0$}
\label{sec:axis}
\begin{figure}
  \centering
    \includegraphics[width=\columnwidth,clip=True]{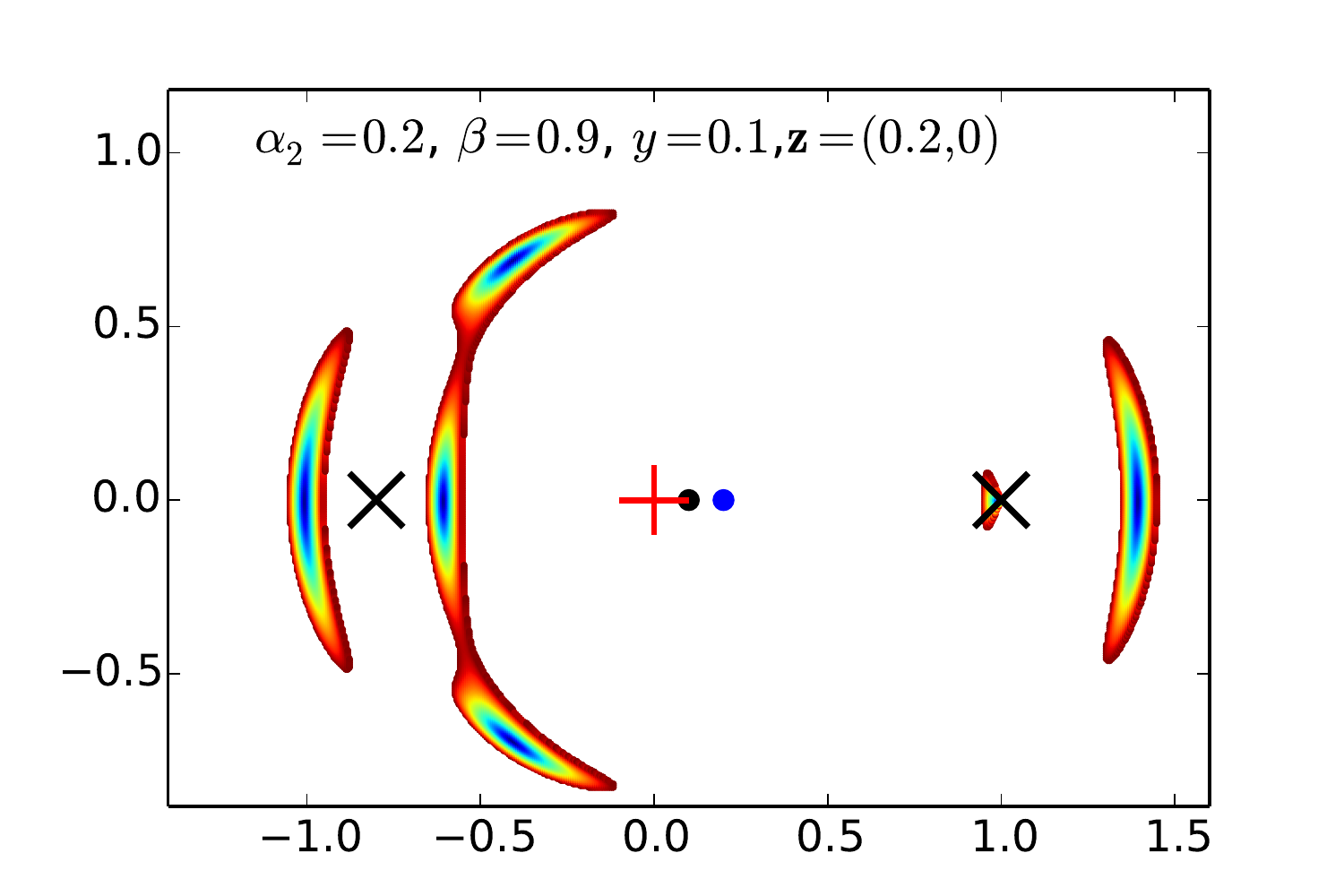}
    \caption {The image plane for a $6^{+2}$ image system. The system has $\alpha_1=1, \alpha_2=0.2,\beta=0.9, y=0.1$ and a circular source centred at $\mathbf{z}=(0.2,0)$, with radius 0.05. The black and blue circles show the unlensed position of the sources on planes 2 and 3. The red cross shows the centre of the first lens, the black crosses show where images of the second lens would form.}
    \label{fig:6images}
\end{figure}

\comments{
\begin{figure*}
  \centering
    \includegraphics[width=0.8\textwidth,clip=True]{alligned.pdf}\\
    \includegraphics[width=0.8\textwidth,clip=True]{beta75.pdf}\\

    \caption {pass}
    \label{fig:alligned}
\end{figure*}
\begin{figure*}
  \centering
  \includegraphics[width=0.8\textwidth,clip=True]{socool.pdf}\\
    \caption {pass}
    \label{fig:socool}
\end{figure*}
}

Simple solutions for $\left(r,\theta\right)$ can be obtained for the case where the lenses and source lie on a plane i.e. $z_j=0$. The $j$ component of Equation \ref{eq:dsplequation2} is thus
\be
\label{eq:zj=0}
0=\left(r-\alpha_1-\frac{\alpha_2(r-\beta \alpha_1)}{D}\right)\sin{\theta}.
\ee
Trivially there are a set of solutions to Equation \ref{eq:dsplequation2} with $\sin{\theta}=0$, and hence $\cos{\theta}=\pm1$ and $x_j=0$. We can thus simplify Equation \ref{eq:dsplequation2} to
\be
z_i=\left(x_i \mp \alpha_1\right)-\frac{\alpha_2}{D}\frac{x_i -y \mp \beta\alpha_1}{|x_i -y \mp \beta\alpha_1|}
\ee
which has up to four real solutions at
\be
x_i=z_i\pm \alpha_1 \pm \alpha_2;
\ee
however the four solutions do not always exist. In turn the conditions are:
\begin{eqnarray}
\label{eqa:conds1}
x_i=z_i + \alpha_1 + \alpha_2 &  \mathrm{exists} \: \mathrm{if} & z_i > y-(\alpha_2+(1-\beta)\alpha_1) \nonumber\\
&  \mathrm{and} & z_i > -(\alpha_1+\alpha_2)\\
\label{eqa:conds2}
x_i=z_i - \alpha_1 - \alpha_2 &  \mathrm{exists} \: \mathrm{if} & z_i < y+(\alpha_2+(1-\beta)\alpha_1) \nonumber\\
&  \mathrm{and} &  z_i < +(\alpha_1+\alpha_2)\\
\label{eqa:conds3}
x_i=z_i - \alpha_1 + \alpha_2 &  \mathrm{exists} \: \mathrm{if} & z_i > y-(\alpha_2-(1-\beta)\alpha_1) \nonumber\\
&  \mathrm{and} & z_i < (\alpha_1-\alpha_2)\\
\label{eqa:conds4}
x_i=z_i + \alpha_1 - \alpha_2 &  \mathrm{exists} \: \mathrm{if} & z_i < y+(\alpha_2-(1-\beta)\alpha_1) \nonumber\\
&  \mathrm{and} & z_i > -(\alpha_1-\alpha_2).
\end{eqnarray}
For the case of $z=0$ and $y=0$, we recover the result of \citet{werner}, that an inner Einstein ring forms only if 
\be
\label{eq:criterion}
1-\beta<\alpha_2/\alpha_1<1.
\ee
{If all four of these images form, two light rays from the source must cross the optical axis between the two lenses, since Equations \ref{eqa:conds3} and \ref{eqa:conds4} require the ray to pass to the left of one lens and the right of the other; we call such systems Einstein zig-zags. Zig-zags form only if Equation \ref{eq:criterion} holds and both
\be
|z-y|< \alpha_2 - \beta \alpha_1 \text{ and } |z|< \alpha_1-\alpha_2
\ee
are satisfied. Rearranging Equations \ref{eqa:conds1} through \ref{eqa:conds4} into constraints on $y$, we recover $|y|< \beta \alpha_1$; for an Einstein zig-zag to form, it is necessary (but not sufficient) that the second lens is multiply imaged by the first. Equation \ref{eq:criterion} is sensible; it requires each lens to be able to refocus a pair of light rays defocused by the other lens. 

Equations \ref{eqa:conds1} to \ref{eqa:conds4} give the location of images forming with $\theta=0 \text{ or } \pi$, i.e. on the plane of the lenses and sources. Images can potentially form off of this plane if Equation \ref{eq:zj=0} can be satisfied for $\sin{\theta}\neq0$.  Equation \ref{eq:zj=0} then implies
\be
\label{eq:D2} 
D=\frac{\alpha_2\left(r-\beta \alpha_1\right)}{r-\alpha_1}
\ee
which can be substituted into the $i$ component of Equation \ref{eq:dsplequation2} to give
\be
z_i=\frac{\alpha_2 y}{D} = \frac{r-\alpha_1}{r- \beta \alpha_1} y;
\ee
hence the images must form at radius
\be
r=\alpha_1\frac{\beta z_i -y}{z_i-y}.
\ee
Curiously, the radius on which the $x_j \neq 0$ images form is independent of $\alpha_2$. 

Substituting $r$ back into Equations \ref{eq:D} and \ref{eq:D2}, we have
\be
\frac{D^2}{y^2}=\left(\frac{\alpha_2}{z_i}\right)^2=\left(\frac{\alpha_1\left(\beta-1\right)}{z_i-y}\right)^2+1-2\left(\frac{\alpha_1\left(\beta-1\right)}{z_i-y}\right)\cos{\theta}
\ee
which gives
\be
\label{eq:conds5}
\cos{\theta}=\frac{1}{2\alpha_1}\left(\alpha_1^2\frac{\beta-1}{z_i-y}+\left(1-\left(\frac{\alpha_2}{z_i}\right)^2\right) \frac{z_i-y}{\beta-1}  \right).
\ee
Since $z,y,\beta,\alpha_1$ and $\alpha_2$, take single values for any lens configuration,  $\cos{\theta}$ has a single value, and hence at most two images can form additionally to those in equations \ref{eqa:conds1} to \ref{eqa:conds4}. For the off-plane images to form, we must have -$1<\cos{\theta}<1$, $r>0$ and $D>0$. The  $r>0$ constraint implies that off-plane images will not form if $\beta z_i<y<z_i$ and the $D>0$ constraint implies off-plane images will not form if $z<0$, but we have been unable to further simplify the inequalities for when the off-axis images do form. 

Equations \ref{eqa:conds1} to \ref{eqa:conds4} and Equation \ref{eq:conds5} show that up to 6 images of the background source can form depending on the values of $z,y,\beta,\alpha_1$ and $\alpha_2$. Forming $6^{+2}$ detectable images only happens in a small range of the parameter space; we show an example case in Figure \ref{fig:6images} - we will call this a $6^{+2}$ image system since there are 6 images of the background source and 2 images of the second lens. In this system we have a lens with two sources at comparable redshift ($\beta=0.9$), the first source is slightly offset from the optical axis ($y=0.1$) and lenses the second source (at $z_i=0.2$) by a tenth of the Einstein radius of the first lens.


\subsection{Numerical solutions for the case $z_j \neq 0$}
\label{sec:offaxis}
\begin{figure}
  \centering
    \includegraphics[width=\columnwidth,clip=True]{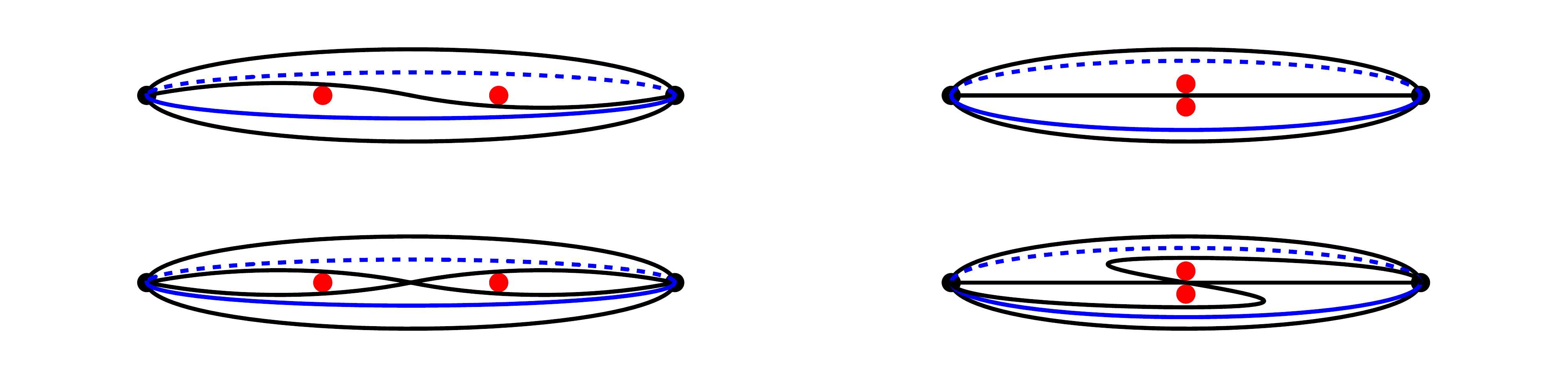}
    \caption {Topological comparison of the compound lens and the binary lens. Possible compound lens topologies, shown on the left, are equivalent to the binary lens topologies on the right. Black lines are light paths in the plane, while blue lines are paths off the plane. The bottom {right} figure is disallowed for strong lensing by a single thin lens plane, since it requires a light ray to double back.}
    \label{fig:topology}
\end{figure}

When the source lies off of the plane described by the observer and the two lenses it is not possible to analytically solve for the image positions\footnote{Rearranging Equation \ref{eq:dsplequation2} into a single equation gives an eighth order polynomial.}; however the image positions can be found numerically. 
 \citet{shin+evans} showed that the caustic structure of the single plane binary lens is complicated and sensitive to the properties of the lenses with up to 7 images forming. Topologically, the compound lens and the binary lens are similar, but the binary lens cannot create the equivalent topology of an Einstein zig-zag (where more than one light ray zig-zags between the two lenses) -- see figure \ref{fig:topology}. The analytical solutions for sources on the $z_j = 0$ line, tell us that zig-zags can only form if Equation \ref{eq:criterion} holds and if the second lens is multiply imaged by the first. For an SIS lens, the zig-zag critical curves collapse to the very centres of the first lens and the two images of the second lens; we can approximate these by circles of very small radius. Lensing these critical curves back to the source plane gives three pseudo-caustics\footnote{A pseudo-caustic is a caustic, but crossing it only creates one detectable additional image, since the second additional image is infinitely demagnified at the core of a singularity.}; if a source is inside all three of these pseudo-caustics, and Equation \ref{eq:criterion} is satisfied, an Einstein zig-zag forms (Figure \ref{fig:caustics1}.1). The critical curves around the images of the second lens map into circles on the source plane with radius equal to the Einstein radius of the second lens and centred on $y \pm (1-\beta) \alpha_1$. The critical curve centred on the first lens forms a pseudo-caustic the shape of which can be found by setting $r \rightarrow 0$ in Equation \ref{eq:dsplequation2}; this pseudo-caustic is shown in red in Figures \ref{fig:caustics1} and \ref{fig:caustics2}. If $\beta \alpha_1 >> y/\alpha_2$, the pseudo-caustic is approximately circular (Figure \ref{fig:caustics1}.1); as $y$ increases the pseudo-caustic becomes kidney bean shape (Figure \ref{fig:caustics1}.3), until the pseudo-caustic loops back on itself to create a limacon (Figure \ref{fig:caustics1}.4). The number of images does not increase if the source is within the loop of the pseudo-caustic compared to being entirely outside the limacon. The lower bound of Equation \ref{eq:criterion} corresponds to when the two circular caustics just touch (Figure \ref{fig:caustics2}.2), and the upper bound corresponds to when the inner and outer loop of the limacon touch for a second time (Figure \ref{fig:caustics2}.5). In Figures \ref{fig:caustics1} and \ref{fig:caustics2} we show the evolution of the limacon caustic as $y$ and $\alpha_2$ vary. If $\alpha_2>\alpha_1$, being inside the limacon destroys an image rather than creating one (Figure \ref{fig:caustics2}.6).

\begin{figure*}
  \centering
    \includegraphics[width=\textwidth,clip=True]{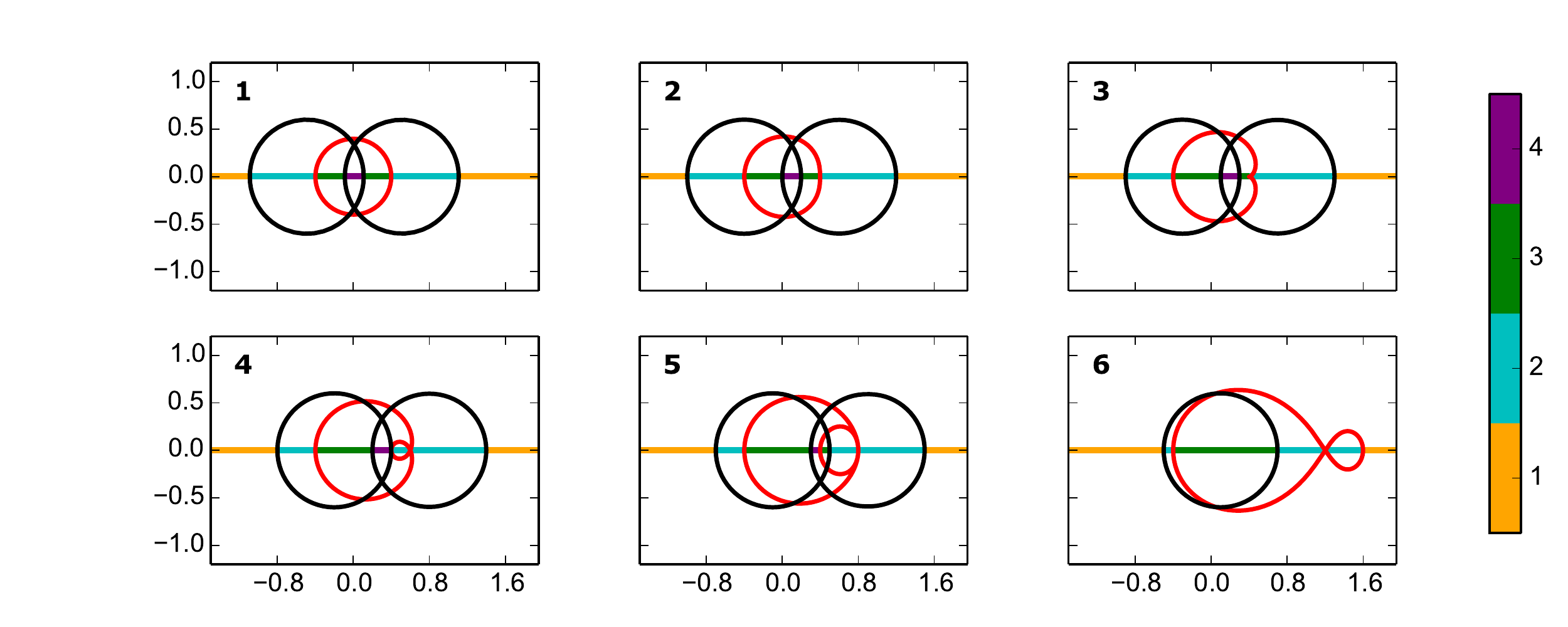}
    \caption {The evolution of the pseudo-caustics of the compound lens as the offset between the two lenses increases. $\beta=0.5$ and $\alpha_2=0.6$ are fixed. The red pseudo-caustic corresponds to the critical curve at the core of the first lens ($r \rightarrow 0$), the circular black pseudo-caustics correspond to the critical curves at the cores of the images of the second lens. The red pseudo-caustic maps to a circle on the second lens plane, but is deformed by the lensing of the second lens. If the second lens is close to the red pseudo-caustic, it is deformed into a limacon on the source plane. }
    \label{fig:caustics1}
\end{figure*}

\begin{figure*}
  \centering
    \includegraphics[width=\textwidth,clip=True]{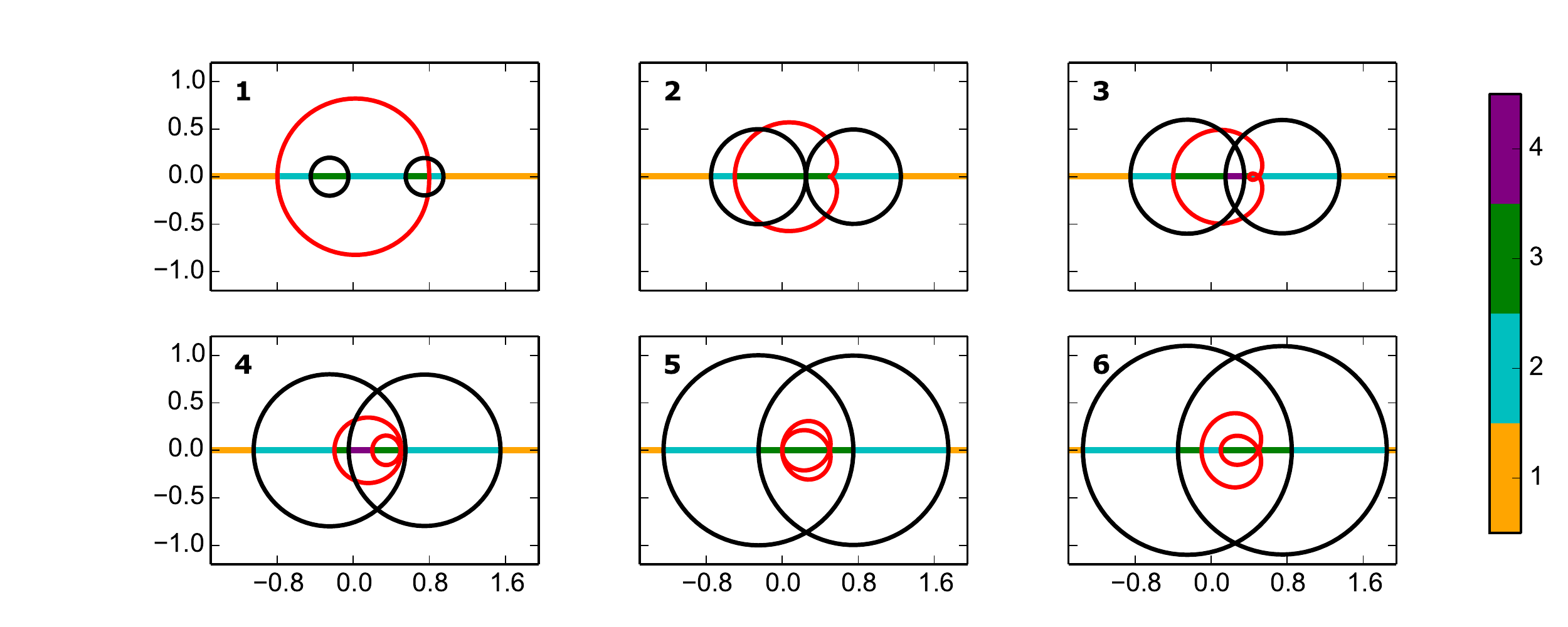}
    \caption {The evolution of the pseudo-caustics of the compound lens as the strength of the second lens increases. $\beta=0.5$ and $y=0.25$ are fixed. The red pseudo-caustic loops back onto itself when $\alpha_2=\alpha_1$ at this point, being inside the red pseudo-caustic, decreases the number of images by one, as rays crossing the optical axis between the first and second lens can no longer be refocused by the first lens.}
    \label{fig:caustics2}
\end{figure*}

\begin{figure*}
  \centering
    \includegraphics[width=0.7\textwidth,clip=True]{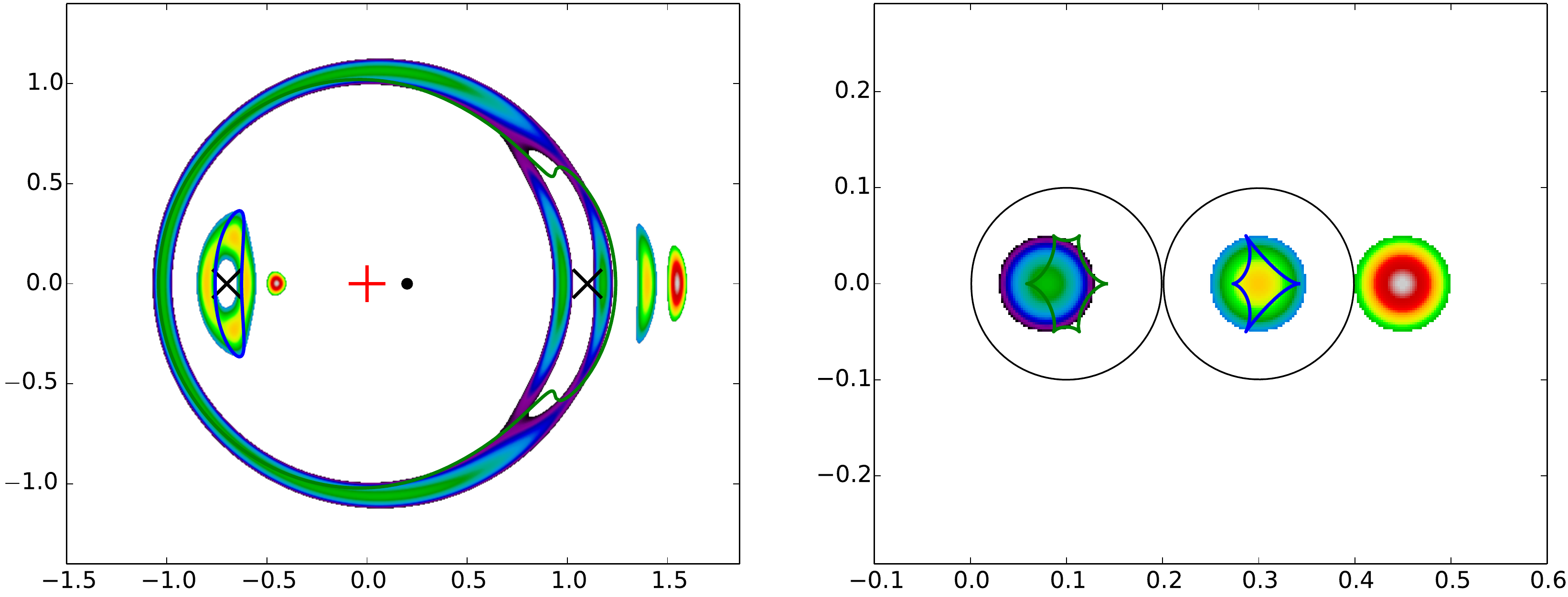}
    \includegraphics[width=0.7\textwidth,clip=True]{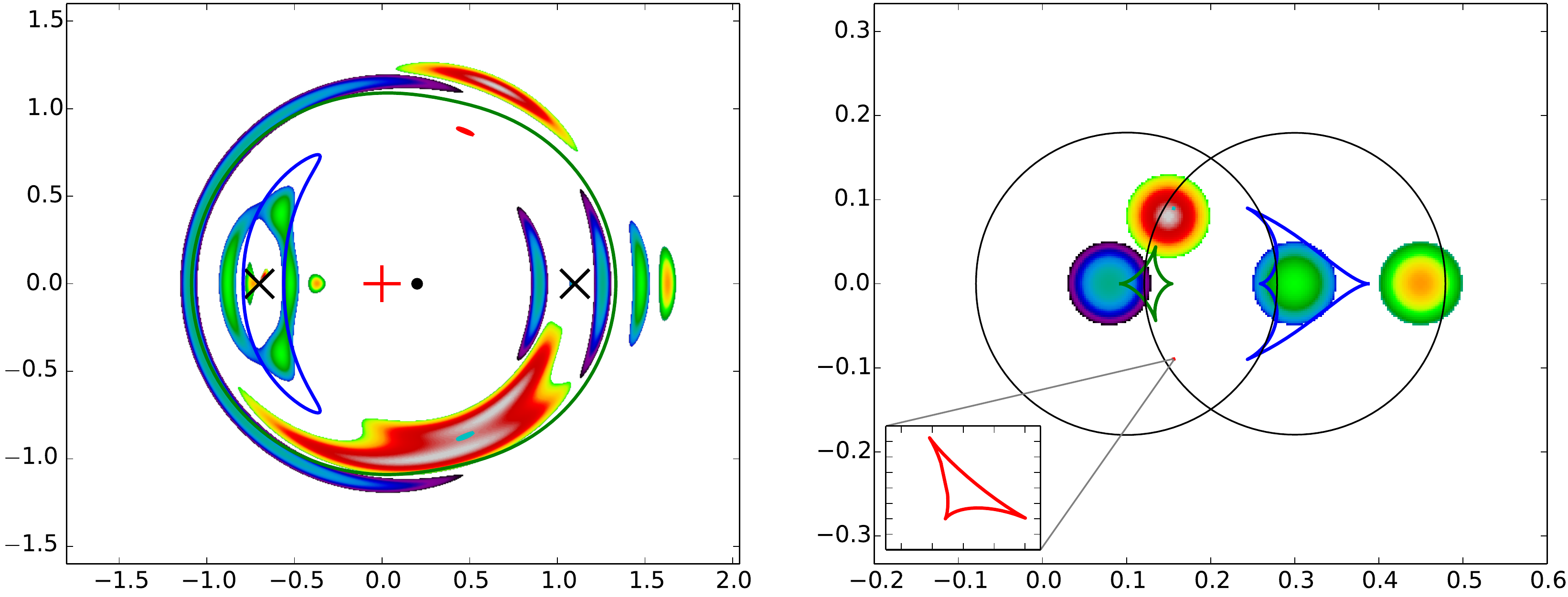}
    \includegraphics[width=0.7\textwidth,clip=True]{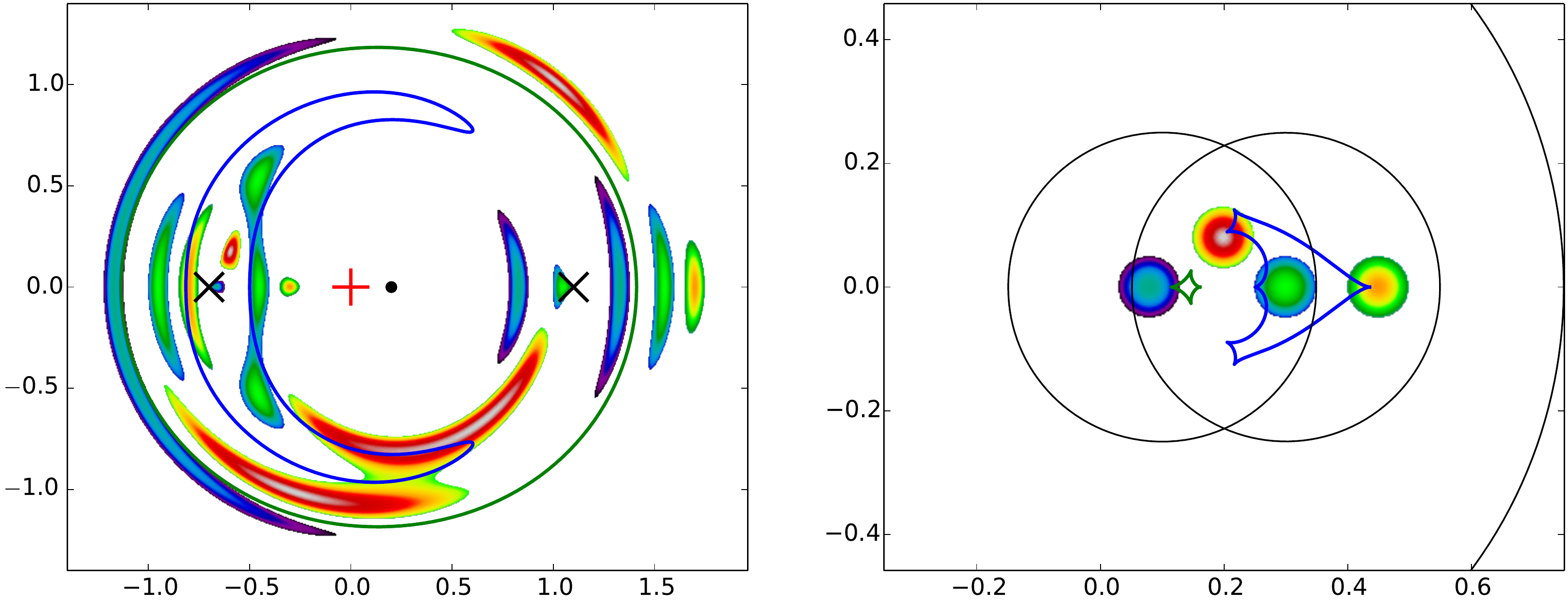}
    \includegraphics[width=0.7\textwidth,clip=True]{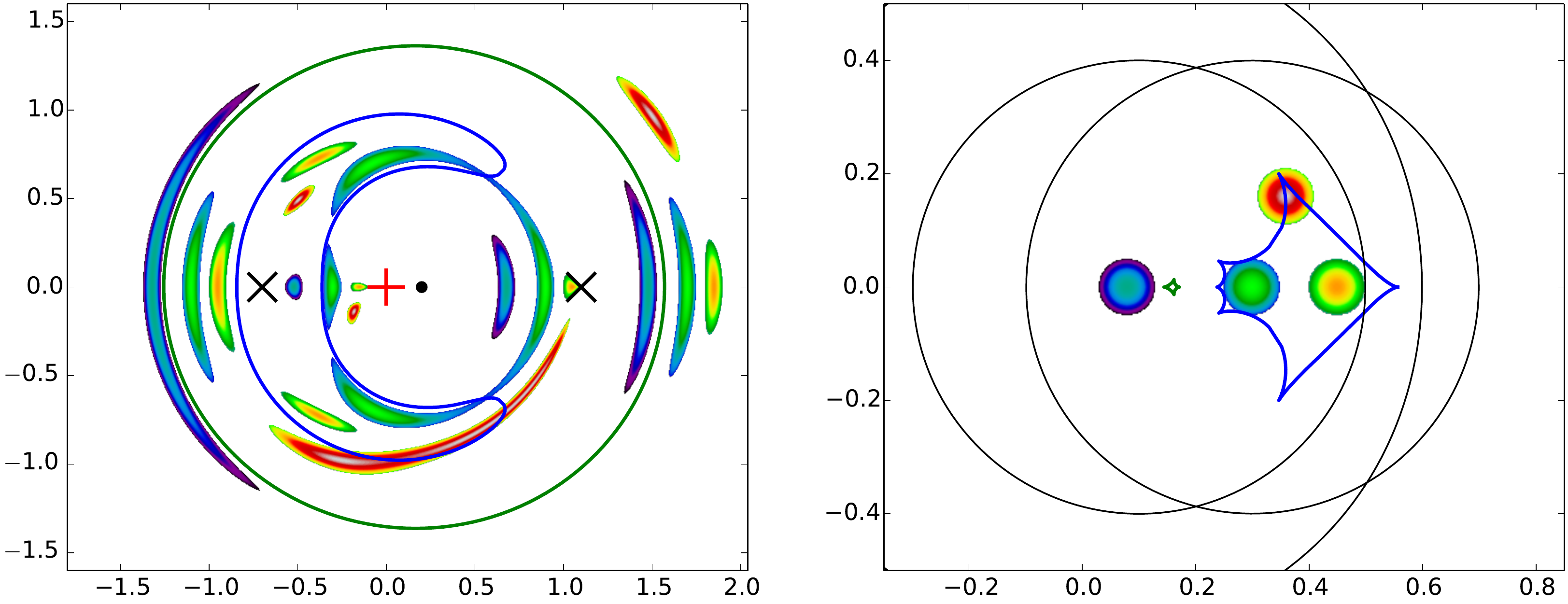}
    \caption {The critical curve {\david (left)} and caustic structure {\david (right)} of a compound lens with $\alpha_1=1, y=0.2$ and $\beta=0.9$. The strength of the second lens increases from top to bottom $\alpha_2 = 0.1, 0.18, 0.25, 0.4$. The black lines are the pseudo-caustics; the pseudo-caustic created by the critical curve at the singularity of the first lens is not always shown on this scale - it is approximately a circle centred at the origin, radius $1-\alpha_2$.}
    \label{fig:images}
\end{figure*}

\begin{figure*}
  \centering
    \includegraphics[width=0.7\textwidth,clip=True]{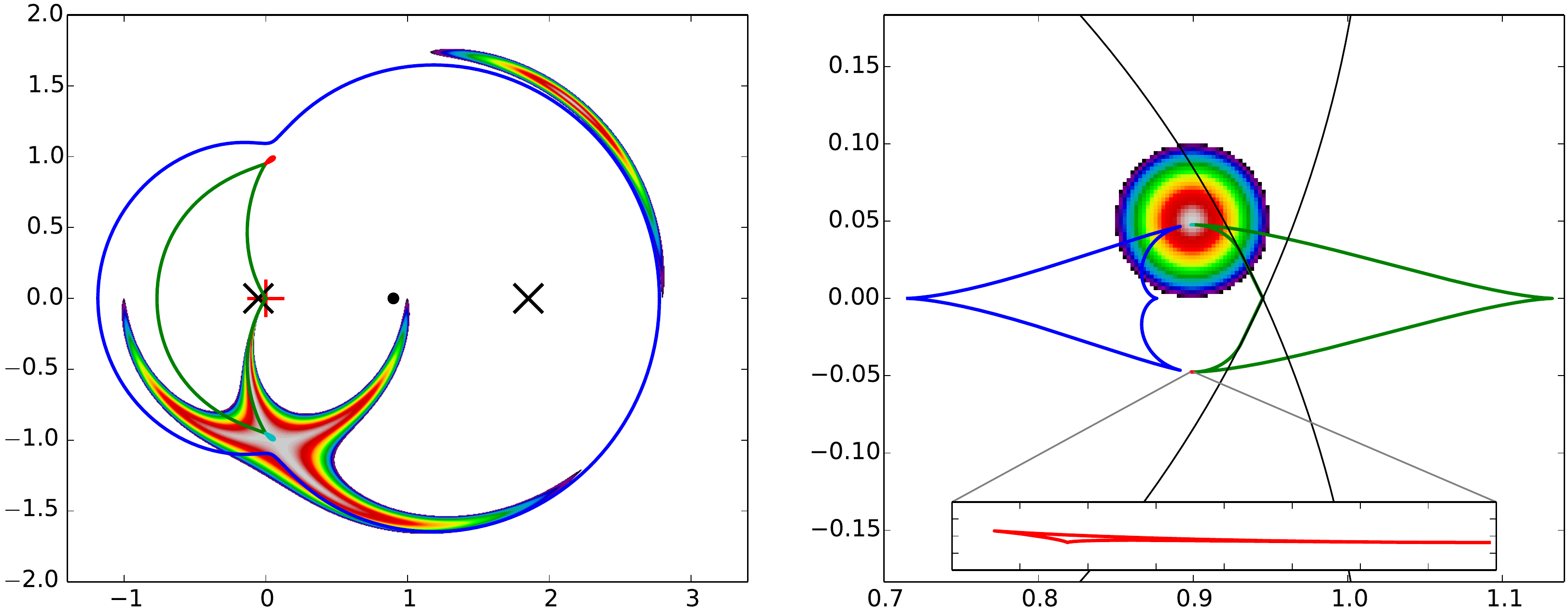}
    \caption {An extremely high magnification event caused by $\alpha_1=1, y=0.9, \alpha_2 = 0.9$ and $\beta^{-1}=1.05$. A small but finite source falls onto the cusps of the blue and green caustics, and covers the tricuspoid caustic, causing very high total magnifications. Such a system could exist for a first lens with  $\sigma_V=190$km/s at $z=0.1$ and a second lens with $\sigma_V=250$km/s at $z=1.6$ acting on a source at $z=10$.}
    \label{fig:images_mental}
\end{figure*}

\begin{figure*}
  \centering
    \includegraphics[width=0.7\textwidth,clip=True]{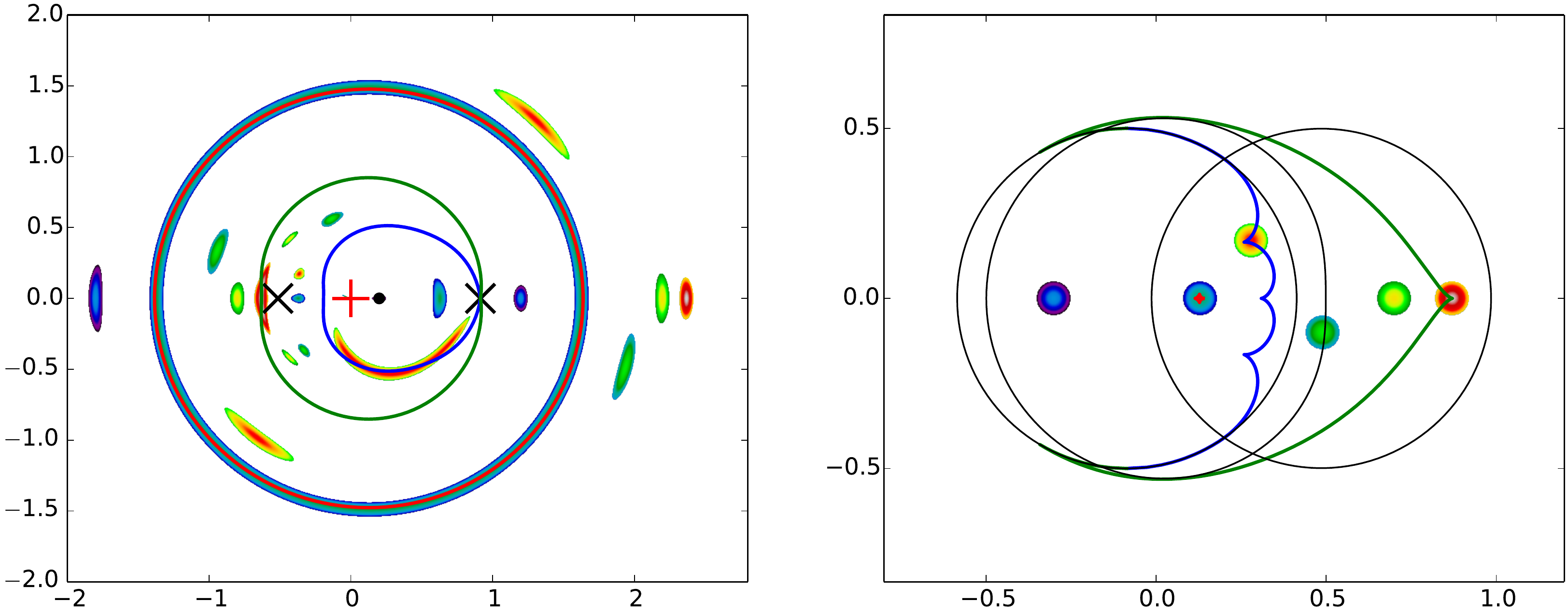}
    \includegraphics[width=0.7\textwidth,clip=True]{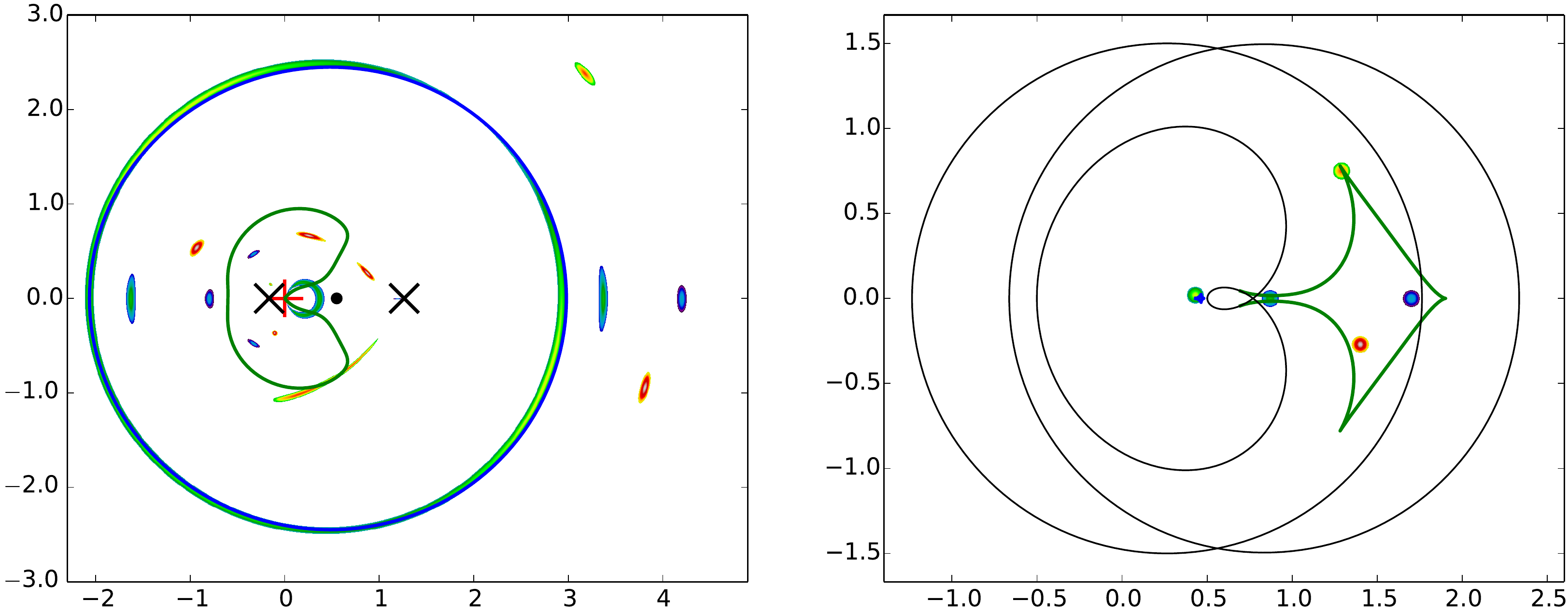}
    \includegraphics[width=0.7\textwidth,clip=True]{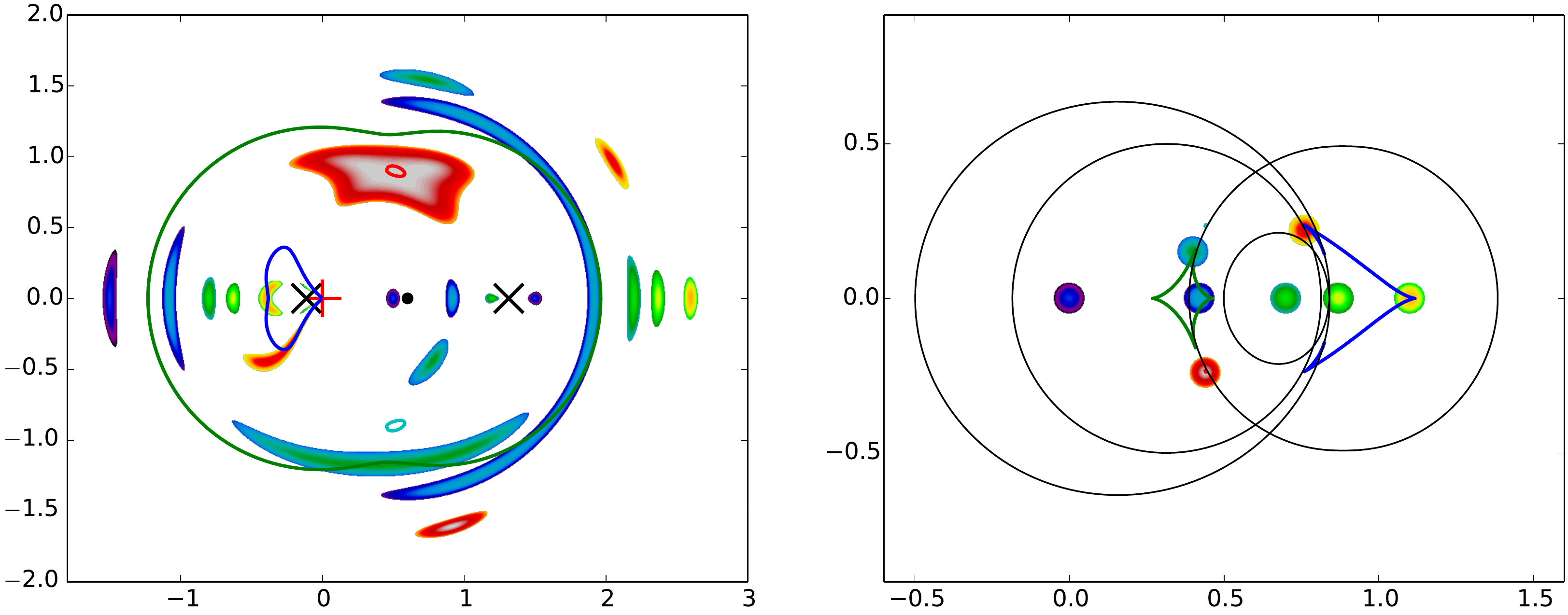}
    \caption {Examples of possible image configurations for three different sets of $\alpha_1, \alpha_2, y, \beta$.}
    \label{fig:images_montage}
\end{figure*}

In addition to the Einstein zig-zag images, analogues of the off-axis images in Section \ref{sec:axis} can also form; again up to 6 images can be detectable. In Figure \ref{fig:images} we show how the critical curves and caustics evolve as $\alpha_2$ increases, holding the other parameters constant. For small $\alpha_2$, two off-axis caustics form; a six cusp ``star'' caustic close to the optical axis and a four-cusp ``Concorde-shaped'' caustic further out. The top panel of Figure \ref{fig:images} shows a source covering the star caustic, which creates an Einstein ring around the primary lens but which bifurcates at the outer image of the second lens; this panel also shows a source covering the Concorde caustic, which creates a smaller Einstein ring around the inner image of the secondary lens. As $\alpha_2$ increases to $\alpha_2>(1-\beta)\alpha_1$ these caustics go through metamorphoses; first a tricuspoid caustic pinches off from the ``star'' caustic (Figure \ref{fig:images}, second panel), before merging into the ``Concorde'' caustic to create a ``bat-shaped'' caustic (Figures \ref{fig:images}, third and fourth panels). The evolution of these caustics are of more than academic interest since the cusps produce extreme magnifications, and caustics with cusps close to each other can produce much higher magnifications of extended sources than is possible with less exotic lenses. This is illustrated in Figure \ref{fig:images_mental}, where a source covers the tricuspoid caustic. A lens configuration this extreme requires some fine tuning of parameters, but a first lens with $\sigma_V=190$km/s at $z=0.1$ and a second lens with $\sigma_V=250$km/s at $z=1.6$ could magnify a $z=10$ source with diameter 1 kpc by a factor of 110, or a source with diameter 0.25 kpc by a factor of 250.

In Figure \ref{fig:images_montage} we show some of the image configurations possible for other plausible compound lens geometries. It is clear from this and Figures \ref{fig:images} and \ref{fig:images_mental}, that the image configurations of many compound lenses are perturbed versions of classic lensing configurations, but for high magnification and high multiplicity events, the image configuration is very different to a typical galaxy-galaxy or galaxy-quasar strong lens.

\section{Beyond two spherical lenses}
\label{sec:2SIE}
Massive galaxies in the real Universe can often deviate significantly from sphericity. Galaxy formation simulations show that the true mass distributions of galaxies are likely to be assymetric and clumpy \citep{vogelsberger}, however models of strong lensing events caused by elliptical galaxies show that elliptically symmetric mass profiles are able to reproduce the observed image positions in most galaxy-galaxy lenses \citep[e.g.][]{sonnenfeld2013,gavazzi2,zuzzana}. If the first lens in a compound system is a singular isothermal ellipsoid it is possible for up to 4 images of the second lens to form, as long as the second lens is close to the optical axis. In Section \ref{sec:offaxis}, we found that for two zig-zag images to form it was necessary that the second lens be doubly imaged. If the second lens is quadruply imaged, one might naively expect it to be possible to create 4 zig-zag images. In fact lensing by two SIE profiles can generate as many as 16 images of a background source as shown in Figure \ref{fig:SIEs}, where a $16^{+4}$ image configuration is generated by an elliptical SIE with a second SIE almost directly behind it. The background source is also almost on the optical axis and the angle between the semimajor axes of the two lenses is only 10 degrees. These high-multiplicity compound lenses often have exotic morphologies; we show a catalogue of some of them in Figure \ref{fig:images_montage}.

\begin{figure}
  \centering
    \includegraphics[width=\columnwidth,clip=True]{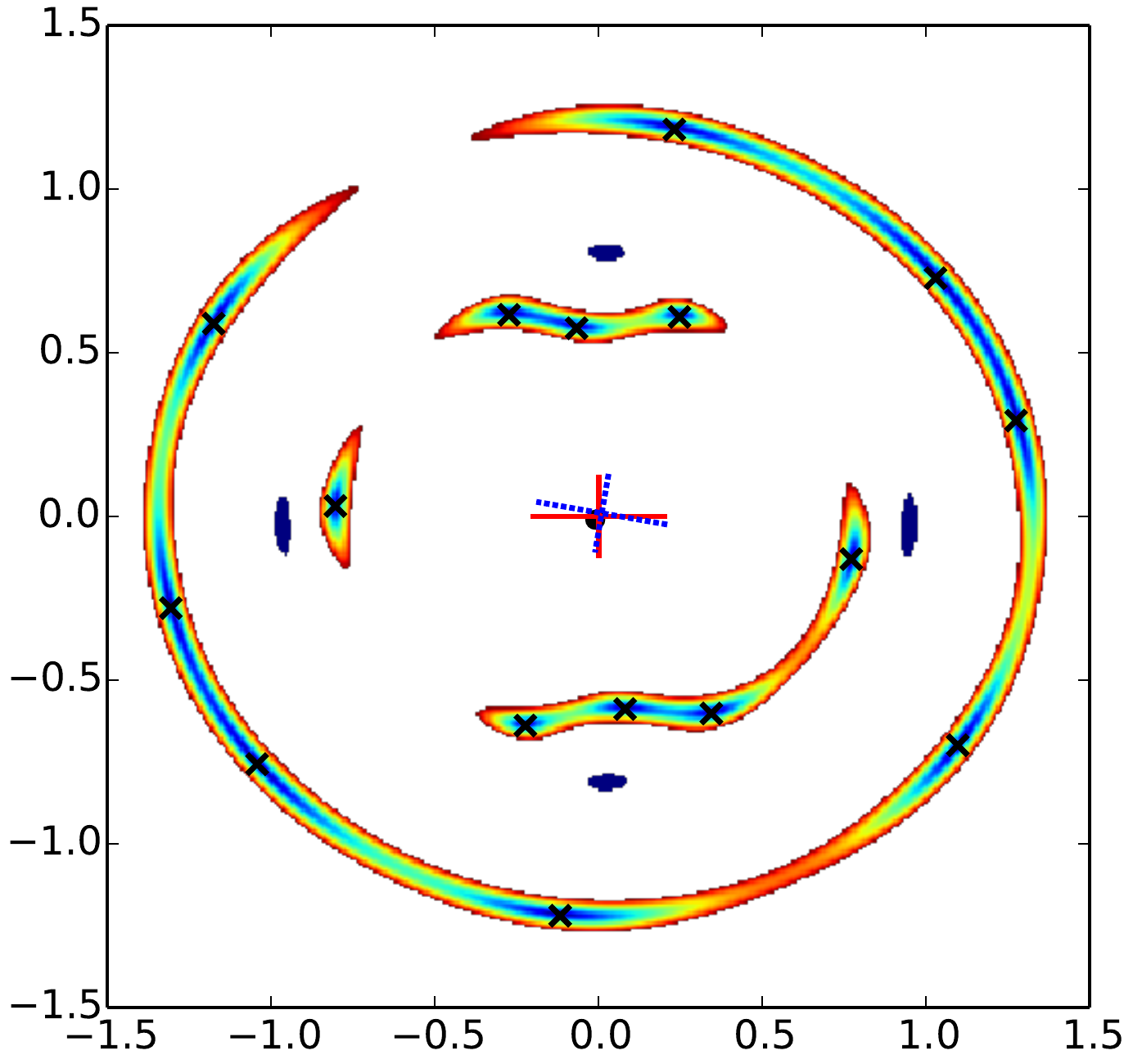}
    \caption {A $16^{+4}$ image configuration caused by two isothermal ellipsoidal mass distributions. Such a high multiplicity arrangement requires lenses and source to lie close to the optical axis, the major-axis of the two lenses to be aligned and the second lens to satisfy equation \ref{eq:criterion}. The 16 image configuration is therefore somewhat contrived. The black crosses show the location that images of a point source would form, while the central region of an extended source is shown in light blue, with the outer regions shown in red. If the second lens is associated with an extended light profile (assumed to be circular, with unlensed radius of 0.05 units), it would appear as the dark blue region. The centre and orientation of the primary lens are shown as the red cross, the unlensed centre and orientation of the second lens are shown as the dashed blue cross, The unlensed position of the background source is shown as a black circle.}
    \label{fig:SIEs}
\end{figure}

\subsection{Compound lensing events in the Universe}
The previous sections have given us an intuitive feel for what compound lensing can produce; however up to now we have made no attempt to discuss how likely different image configurations are. To assess how common different compound lensing events are, we must first understand the populations of deflectors in the Universe. \citet{collett2015a} investigated the frequency of strong lensing events in forthcoming surveys. We use the {\sc{LensPop}} code of \citet{collett2015a} to generate our population of deflectors. The first deflector is drawn from the observed distribution of velocity dispersions and ellipticities observed in the Sloan Digital Sky Survey \citepeg{choi, collett2015a}, with deflectors assumed to be randomly distributed in a co-moving volume. The {\sc{LensPop}} source catalogue is based upon semi-analytic models that paint sources onto the Millennium Simulation \citep[see][for details]{connolly,delucia}, with number counts re-scaled to match observations. These catalogues give stellar masses that we convert into velocity dispersions using the relation of \citet{slacsX}; $\log_{10}(\sigma/\text{km/s})=0.18 \times\log_{10}(M_*/10^{11} M_{\odot})+2.34)$. We assume that the mass profile has the same alignment and flattening as the light profile. With this model we can ray-trace back to assess how common it is for compound lenses to produce high multiplicity images or extreme magnifications.

For computational speed and to ensure that the systems are at least plausibly resolved by current and next-generation telescopes, we neglect lenses that have $\beta\theta_{E1} <0.2"$, $\theta_{E1} <0.5"$ or $\theta_{E2} <0.1"$ (where the Einstein radii are defined in terms of the final source plane). We do not count images that have an (unsigned) magnification less than 0.1, as such demagnified images are likely to be hard to detect in practice. We count groups of multiple images separated by less than 0.1 arcseconds as one image when assessing image multiplicity, as the scientific utility of such a group of images is likely to be the same as a single image, given the resolving power of current telescopes.

We find that the optical depth for compound lensing is roughly $6\times 10^{-6}$  for $z>3$. High multiplicity images are rarer, but of order  $4\times 10^{-8}$  of the $z>3$ sky is compound lensed to produce 6 or more images. The optical depth for multiple imaging is shown as a function of redshift in Figure \ref{fig:tauN}. We find that the optical depth for compound lensing increases by an order of magnitude between $z=1$ and $z=2$ but is broadly flat between $z=4$ and $z=10$. Once magnification bias is included, we may expect to find such events in billion-galaxy surveys.

The optical depth for magnification by a compound lens is shown as a function of redshift in Figure \ref{fig:tauM}. For extreme magnifications we find that for sources with a 1 kpc radius compound lensing causes $\sim$10$^{-6}$ of sources at $z>2$ to be increased by two or more magnitudes. Excitingly a $z=10$ source with radius 0.1 kpc will be magnified by 4 magnitudes $\sim$10$^{-7}$ of the time, due to compound galaxy-galaxy lensing. Due to the exponential cutoff of the high redshift luminosity function, the population of high-z sources discovered in future surveys may be significantly biased towards extreme magnification events. Assuming that high-z sources are 0.1 kpc radius discs, and that the Euclid detection threshold for compound lenses is the same as for extended sources, the luminosity function of \citet{masonLF} predicts $1.0_{-0.4}^{+0.5}$ redshift 8 source that is magnified by 100 or more will be detectable with Euclid (C. Mason, private communication)

\begin{figure}
  \centering
    \includegraphics[width=\columnwidth,clip=True]{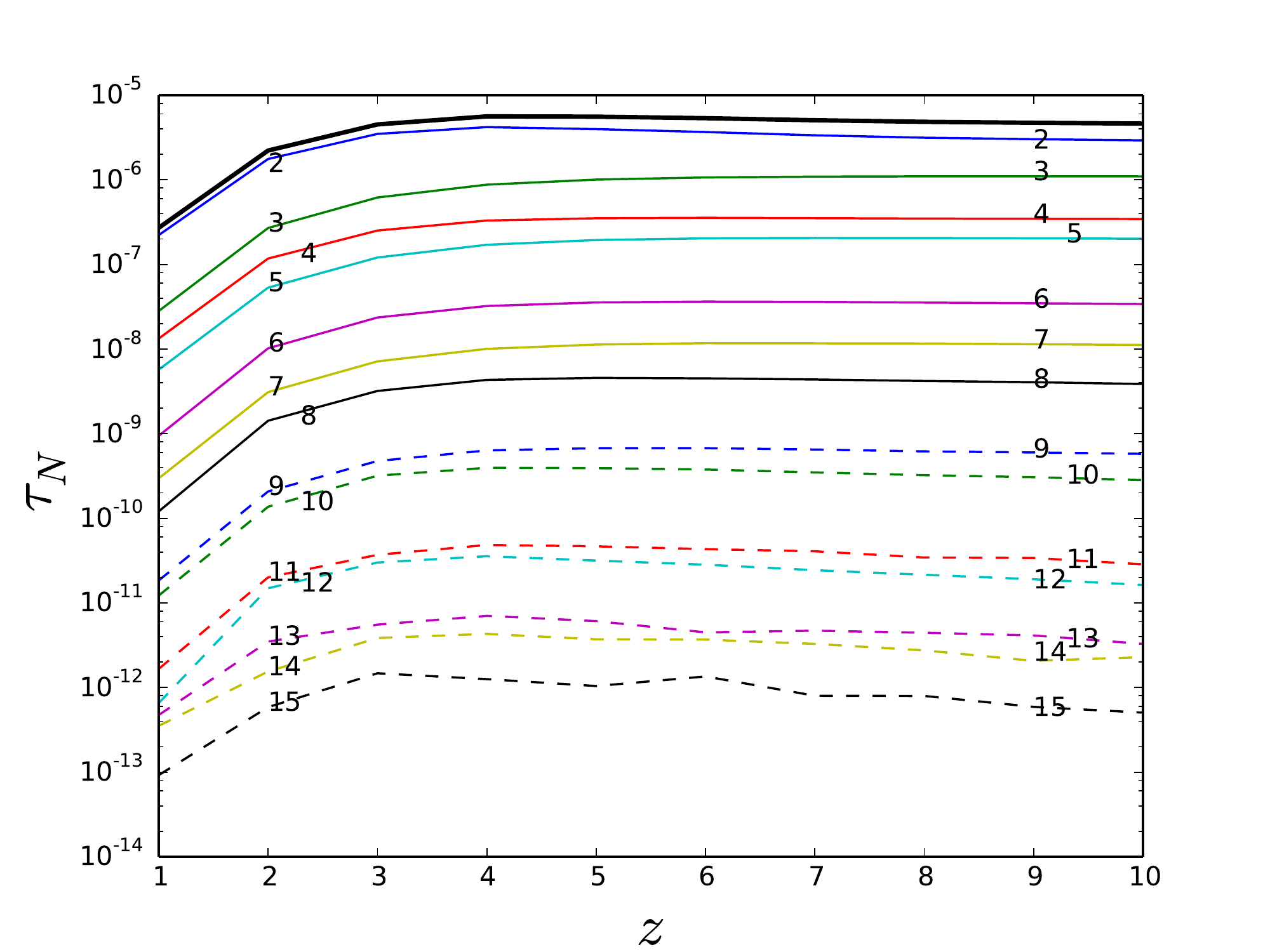}
    \caption {The optical depth for multiple imaging by a compound lens as a function of source redshift. The black solid line represents the total optical depth for multiple imaging by a compound lens. Each coloured line represents a specific multiplicity of imaging as denoted by the number on the line.}
    \label{fig:tauN}
\end{figure}

\begin{figure}
  \centering
    \includegraphics[width=\columnwidth,clip=True]{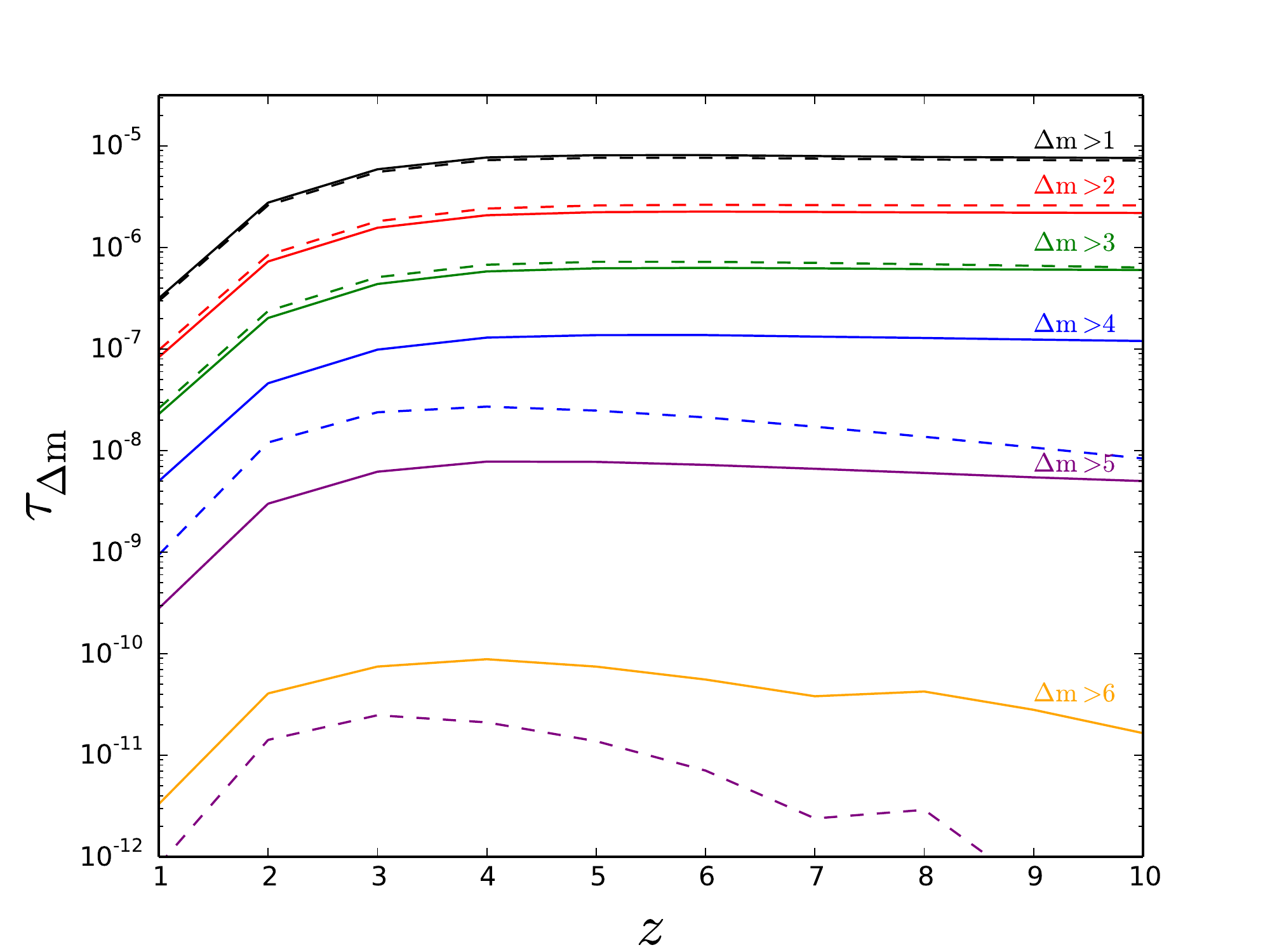}
    \caption {{The optical depth for magnification causing changes in the apparent magnitude ($\Delta$m) of an extended source by a compound lens as a function of source redshift.} The solid (dashed) lines are for a uniform circular source with radius 0.1 kpc (1 kpc).}
    \label{fig:tauM}
\end{figure}
\section{Conclusion}
\label{sec:conclude}

In this paper we have investigated compound lensing by two lens galaxies and a further source, at different redshifts, and the image morphologies this can generate. We found that compound lensing can produce high multiplicity and high magnification events. Discovering a high multiplicity lensed quasar would be an extremely powerful cosmological probe, since many independent time-delays can be measured. High magnification events have the power to uncover the population of ultra-faint high redshifts sources that are otherwise far beyond the depth and resolution of current telescopes.

We found analytic results for the case of two SIS lenses, solving completely for the case when the observer, source and both lenses lie on a plane. From this we found that the highest multiplicity images can only form if both (a) the second lens is multiply imaged by the first lens, and (b) the Einstein radius of the second lens is comparable to, but less than the Einstein radius of the first lens. We showed numerically that when the source lies off of the plane, some lens configurations can generate higher order catastrophes that produce extreme magnifications - a potentially important method for discovering and studying intrinsically faint high redshift sources. The simple picture painted by the two SIS model is complicated significantly when the lenses are elliptically flattened. Up to 16 image configurations can exist if the second lens is quadruply imaged by the first lens. 

With this analytic understanding of the system, we then used the {\sc LensPop} code of \citet{collett2015a} to answer our three primary questions:

\begin{itemize}
  \item How often does compound lensing generate high multiplicity systems?
\end{itemize}
We found that the optical depth for multiple imaging by a compound lens is $\simeq 6 \times 10^{-6}$  for $3<z<10$. Image multiplicities of 6 or greater are 2.5 orders of magnitude rarer. However once magnification is included, a handful may be discovered in a wide-deep survey like LSST which should detect $\sim 2 \times 10^{10}$ galaxies \citep{lsst}.

\begin{itemize}
  \item How often does compound lensing generate extreme magnifications?
\end{itemize}
We found $\simeq 2 \times 10^{-6}$  of the  $3<z<10$ sky has its flux increased by more than two magnitudes. Magnification of 100 (decreasing the apparent magnitude by 5) is 2.5 orders of magnitude rarer for 0.1 kpc radius sources, but is almost impossible for sources with 1 kpc radii. Under simplifying assumptions, we inferred that Euclid may discover one $z=8$ source that has been magnified  by a hundred.

\begin{itemize}
  \item What do high multiplicity/magnification compound lenses look like?
\end{itemize}
For low multiplicity/magnification compound lenses, the image configurations are typically perturbed versions of single-plane image morphologies. However, for high multiplicity/magnification compound lenses, the image morphologies are significantly more complicated; even with two SIE profiles the parameter space is rich, and many images can form. We show 20 high multiplicity configurations in Figure \ref{fig:images_montage}, and {\david there is a real danger that current lens finding algorithms would fail to detect any of these as lens candidates}. The most exotic lenses will require novel discovery algorithms, or an extensive and imaginative human search.


\begin{figure*}
  \centering
    \includegraphics[width=0.24\textwidth,clip=True]{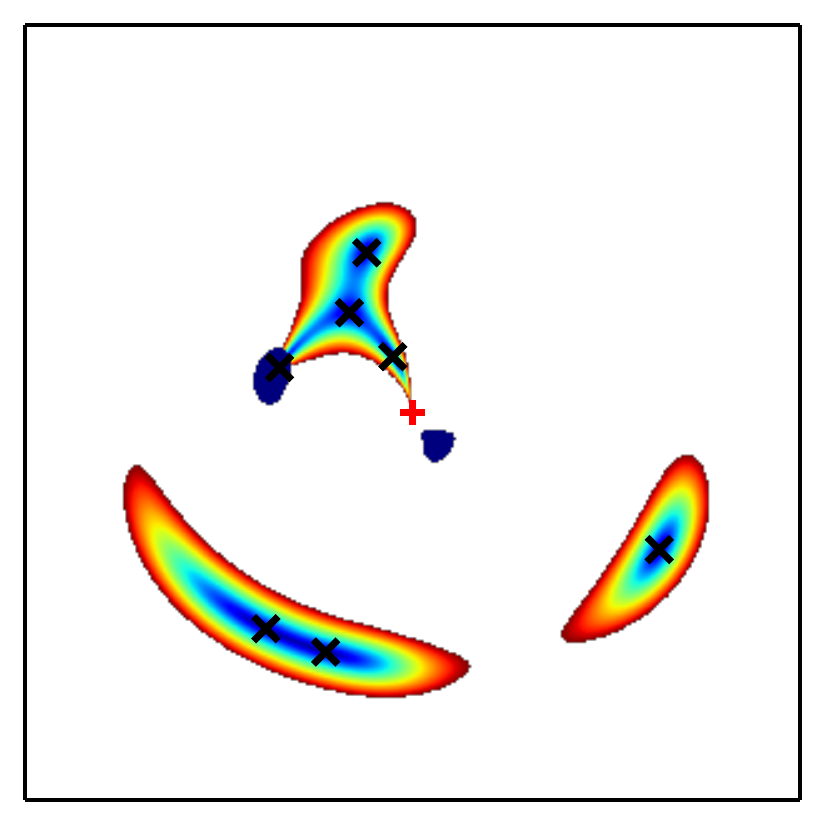}
    \includegraphics[width=0.24\textwidth,clip=True]{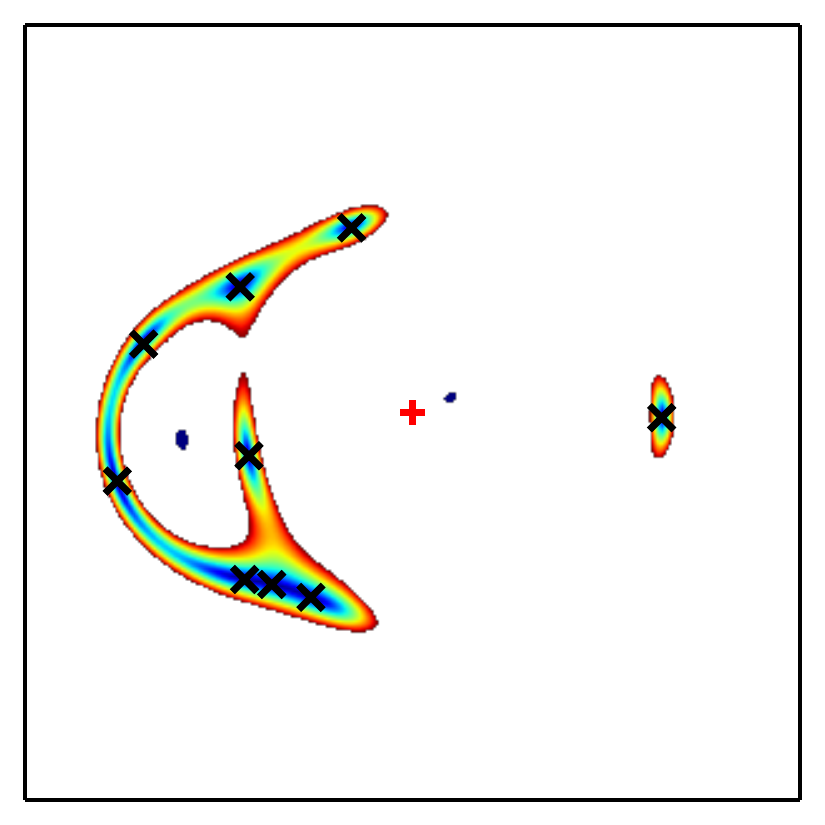}
    \includegraphics[width=0.24\textwidth,clip=True]{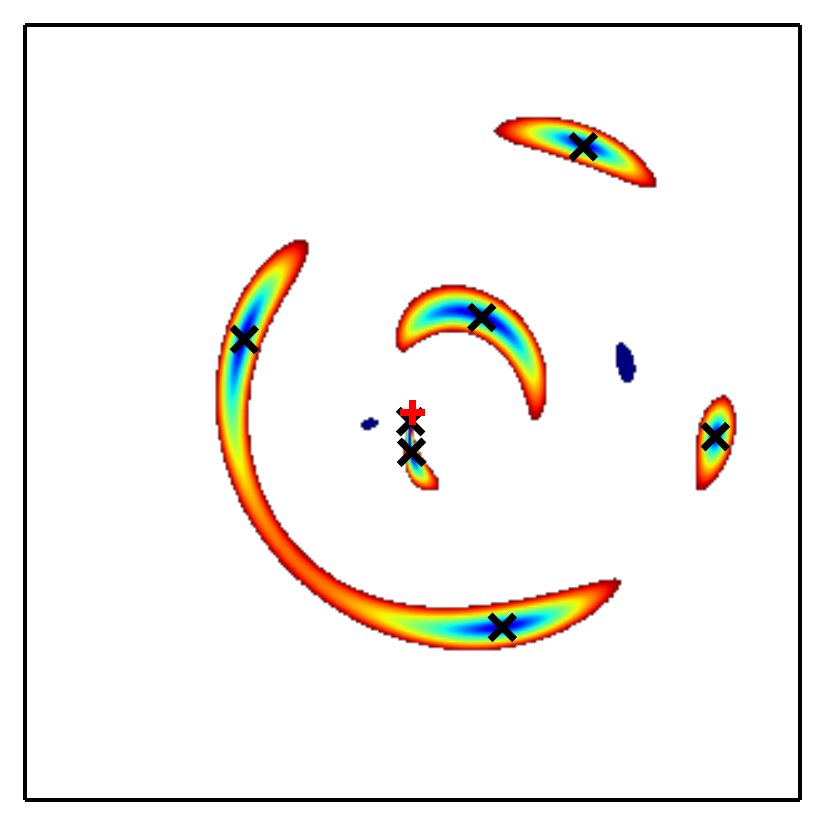}
    \includegraphics[width=0.24\textwidth,clip=True]{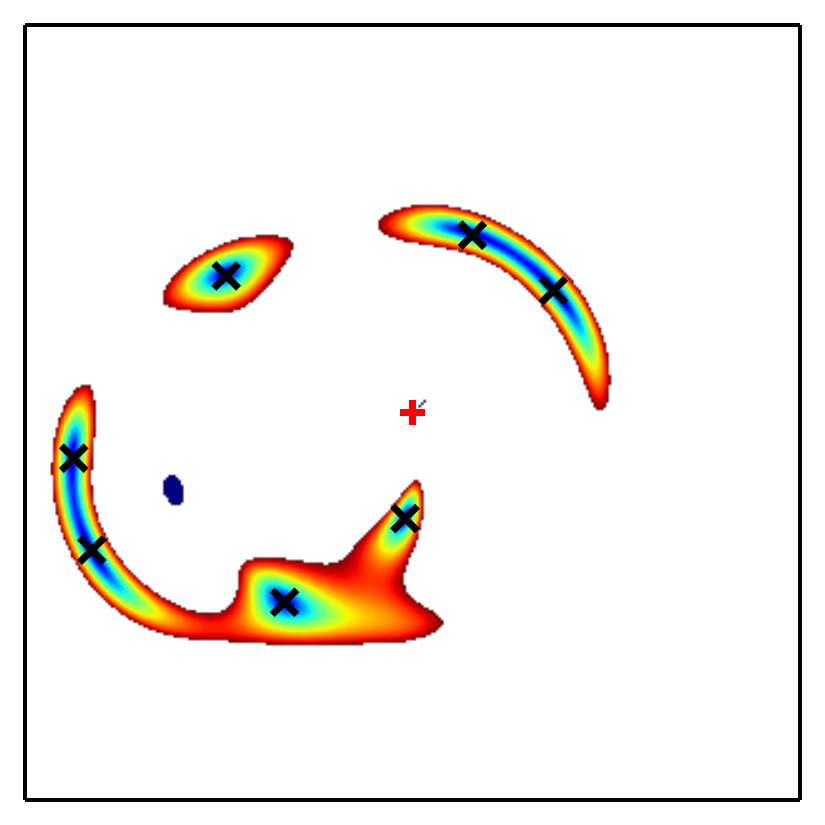}
\\
    \includegraphics[width=0.24\textwidth,clip=True]{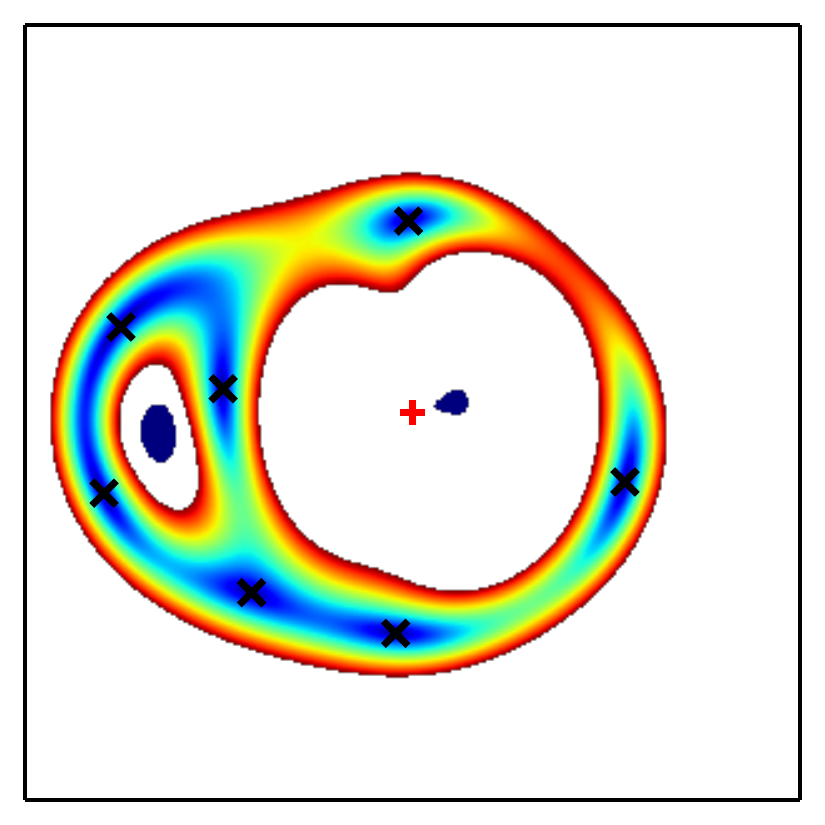}
    \includegraphics[width=0.24\textwidth,clip=True]{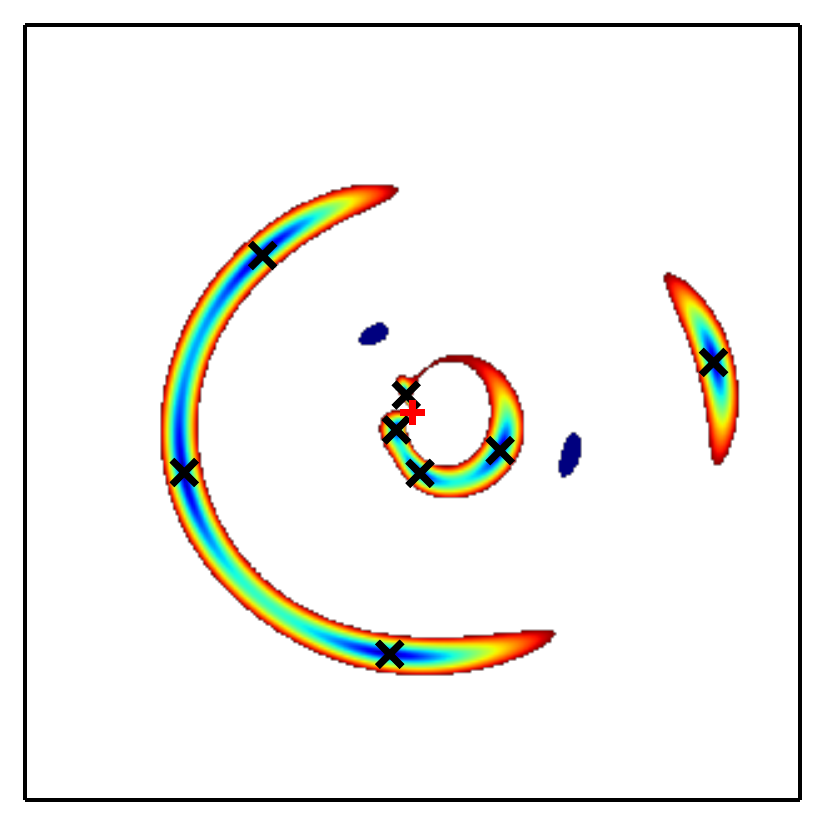}
    \includegraphics[width=0.24\textwidth,clip=True]{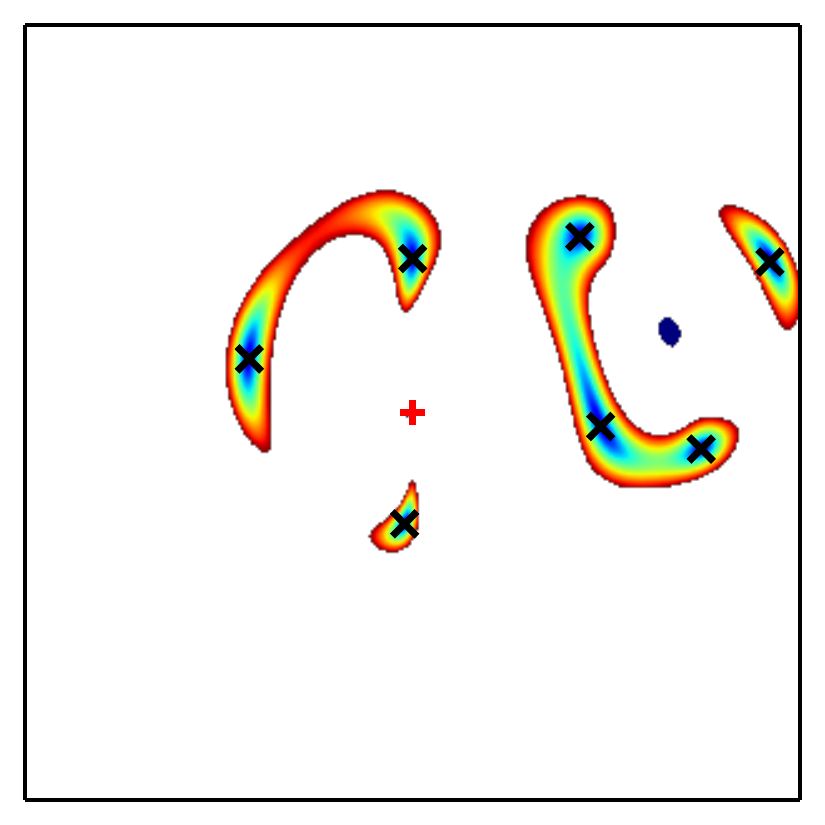}
    \includegraphics[width=0.24\textwidth,clip=True]{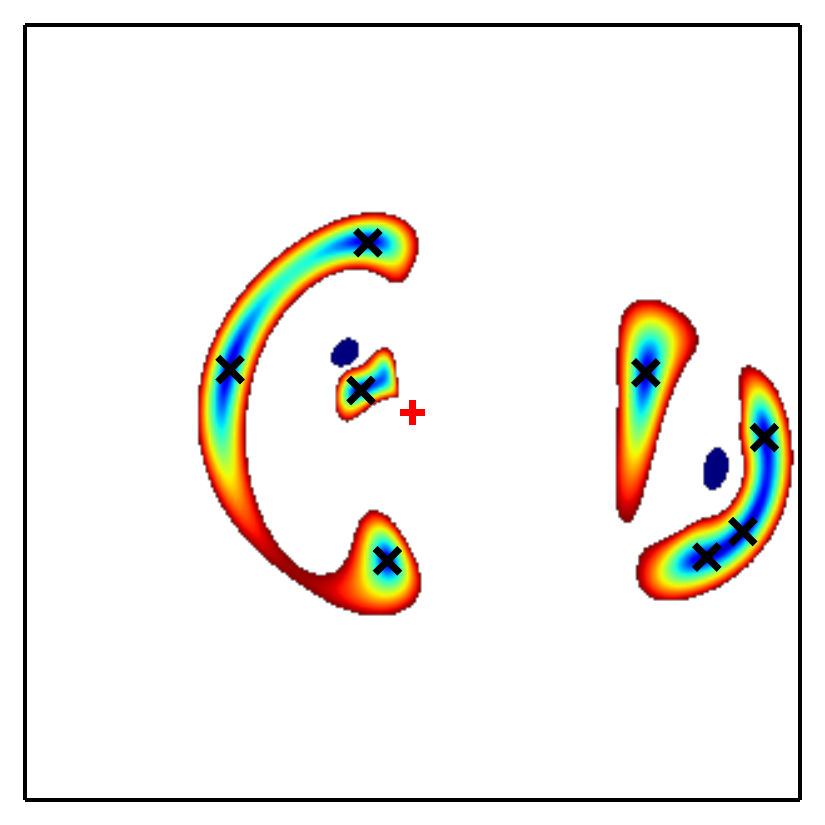}
\\
    \includegraphics[width=0.24\textwidth,clip=True]{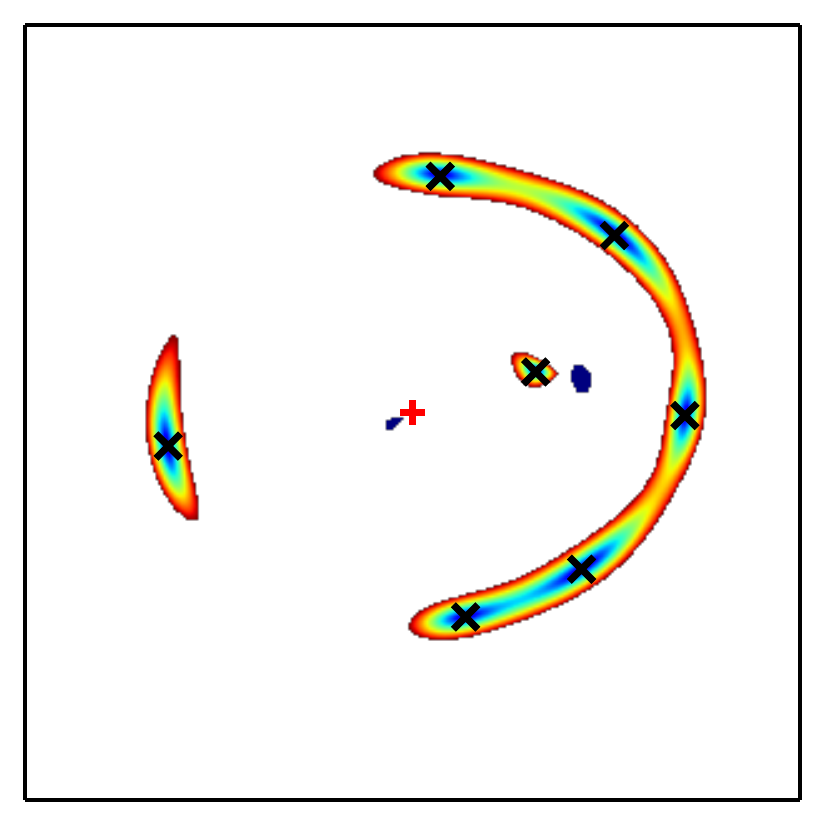}
    \includegraphics[width=0.24\textwidth,clip=True]{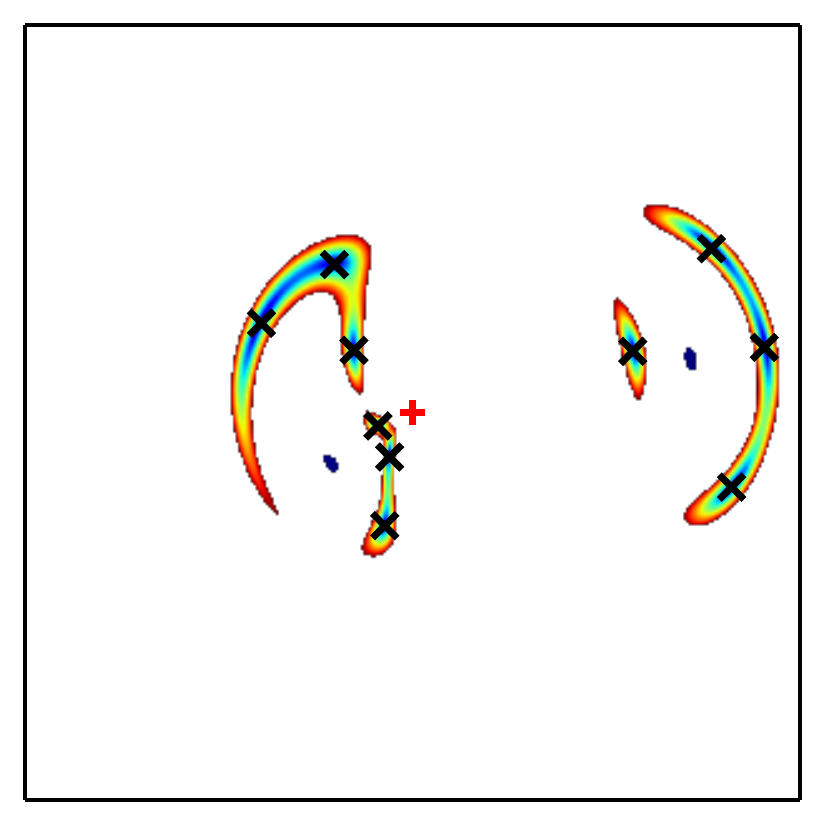}
    \includegraphics[width=0.24\textwidth,clip=True]{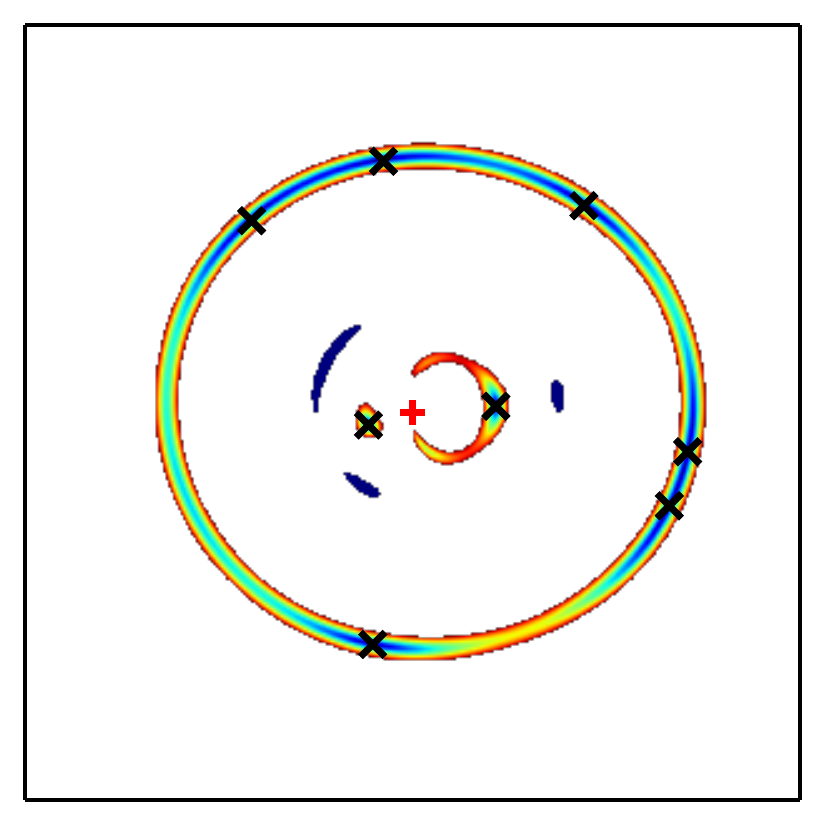}
    \includegraphics[width=0.24\textwidth,clip=True]{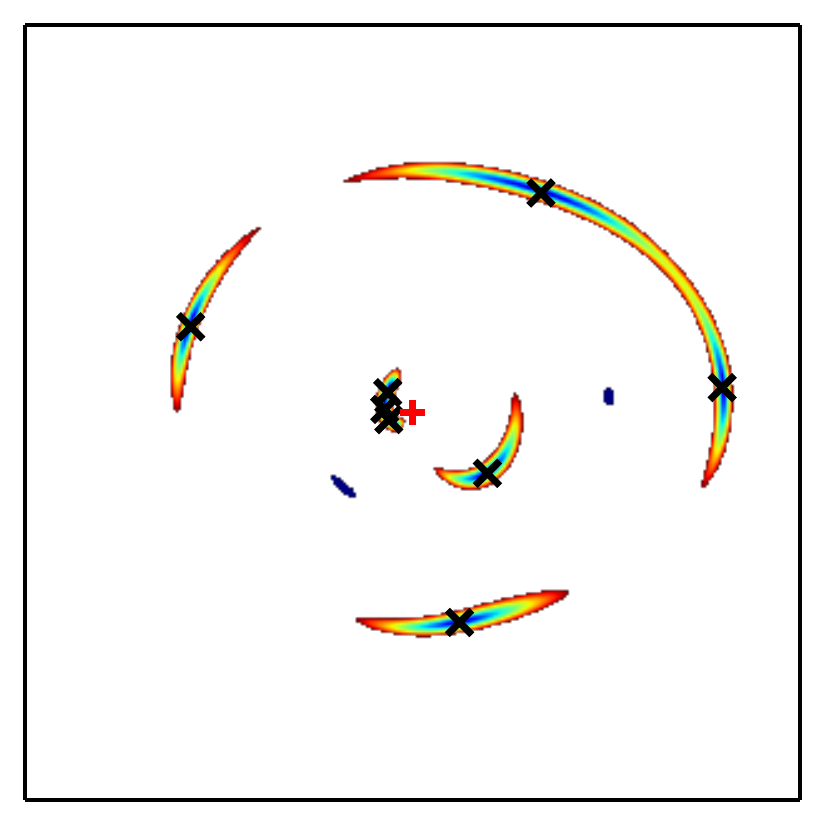}
\\
    \includegraphics[width=0.24\textwidth,clip=True]{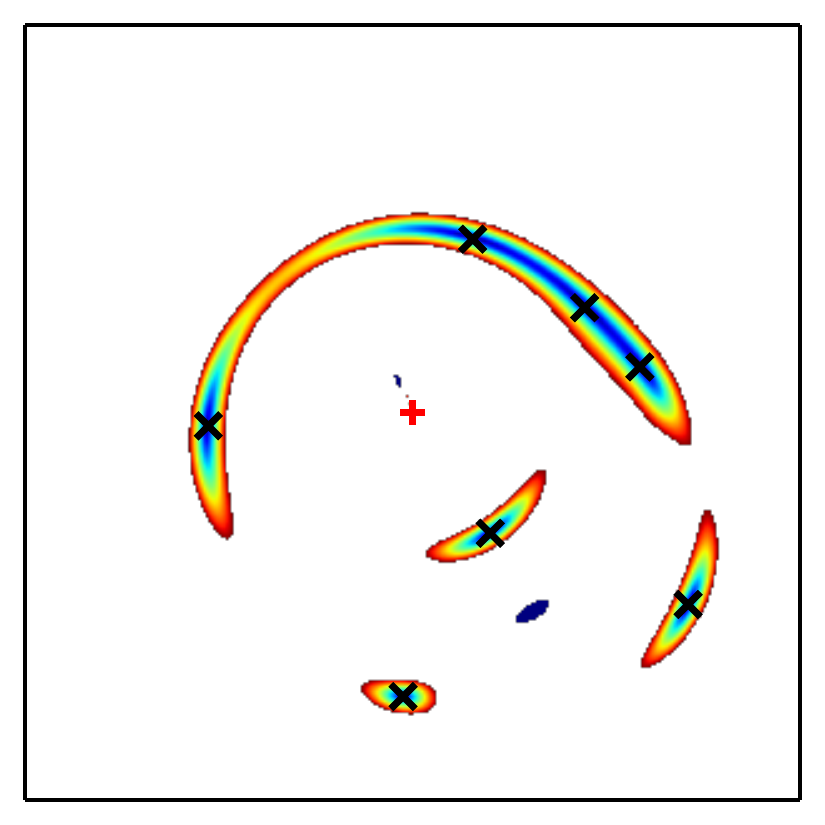}
    \includegraphics[width=0.24\textwidth,clip=True]{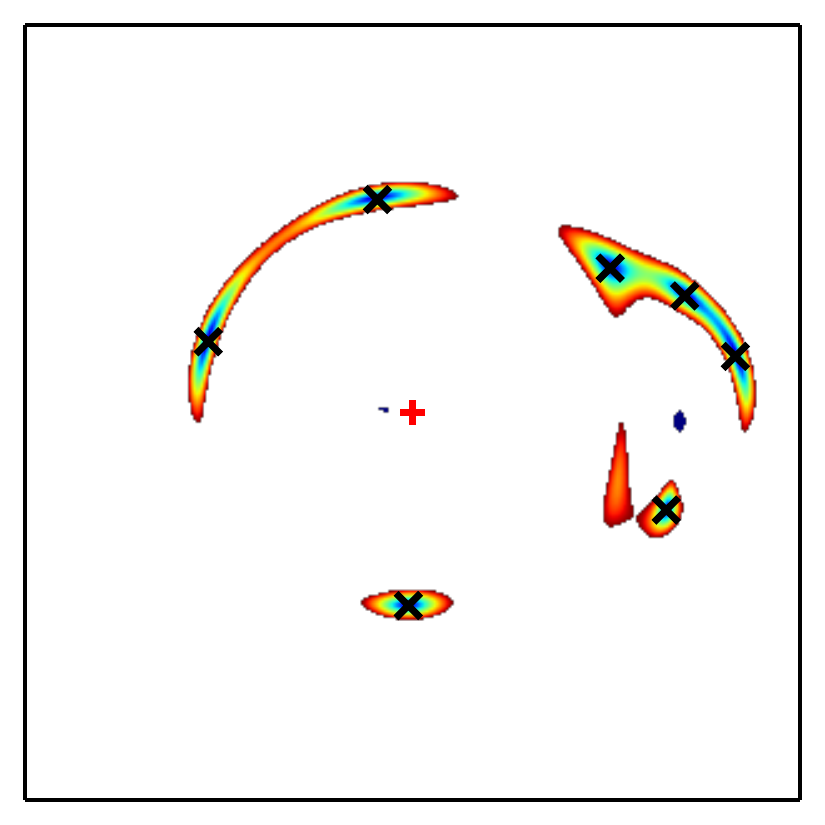}
    \includegraphics[width=0.24\textwidth,clip=True]{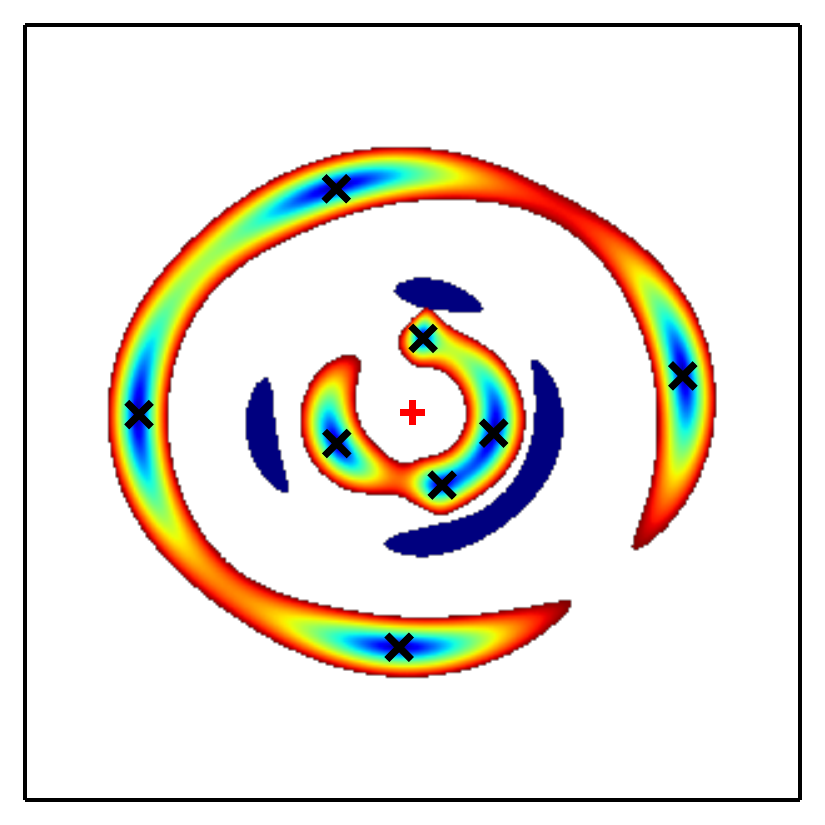}
    \includegraphics[width=0.24\textwidth,clip=True]{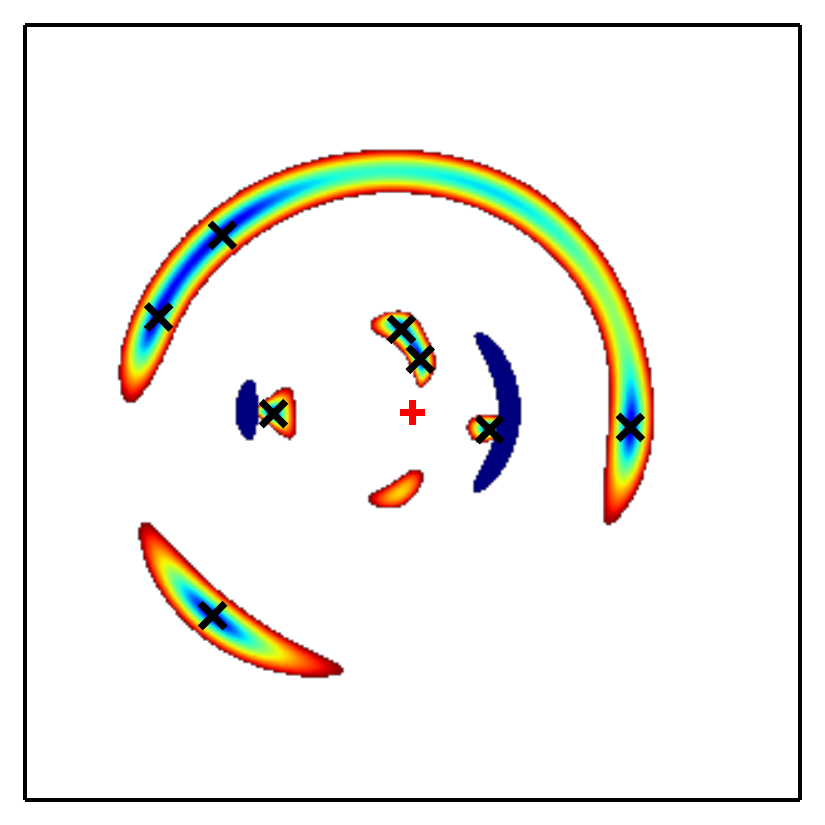}
\\
    \includegraphics[width=0.24\textwidth,clip=True]{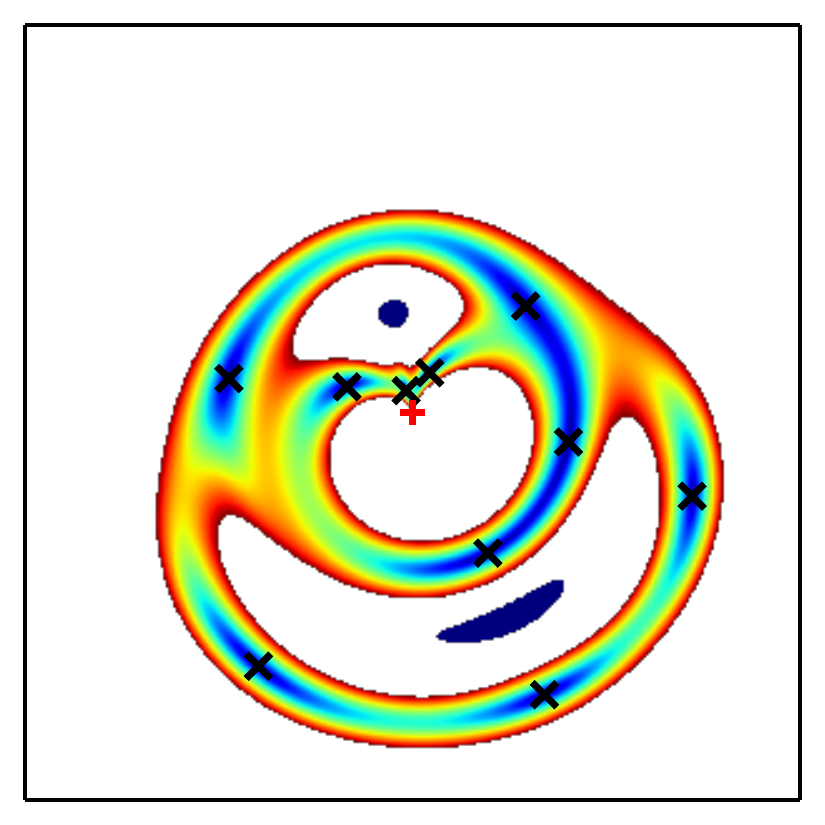}
    \includegraphics[width=0.24\textwidth,clip=True]{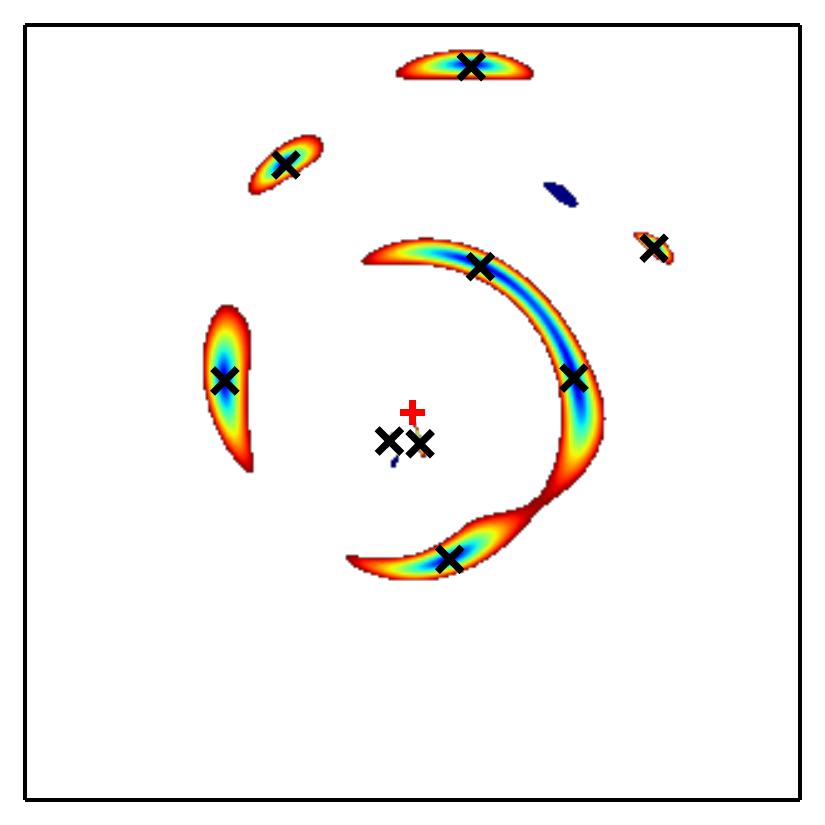}
    \includegraphics[width=0.24\textwidth,clip=True]{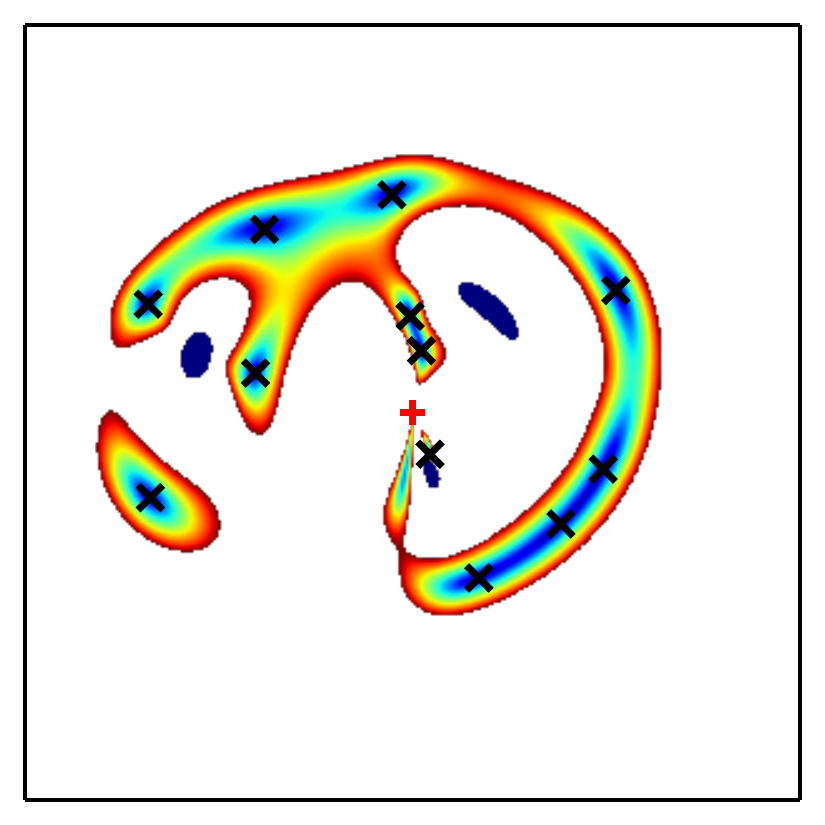}
    \includegraphics[width=0.24\textwidth,clip=True]{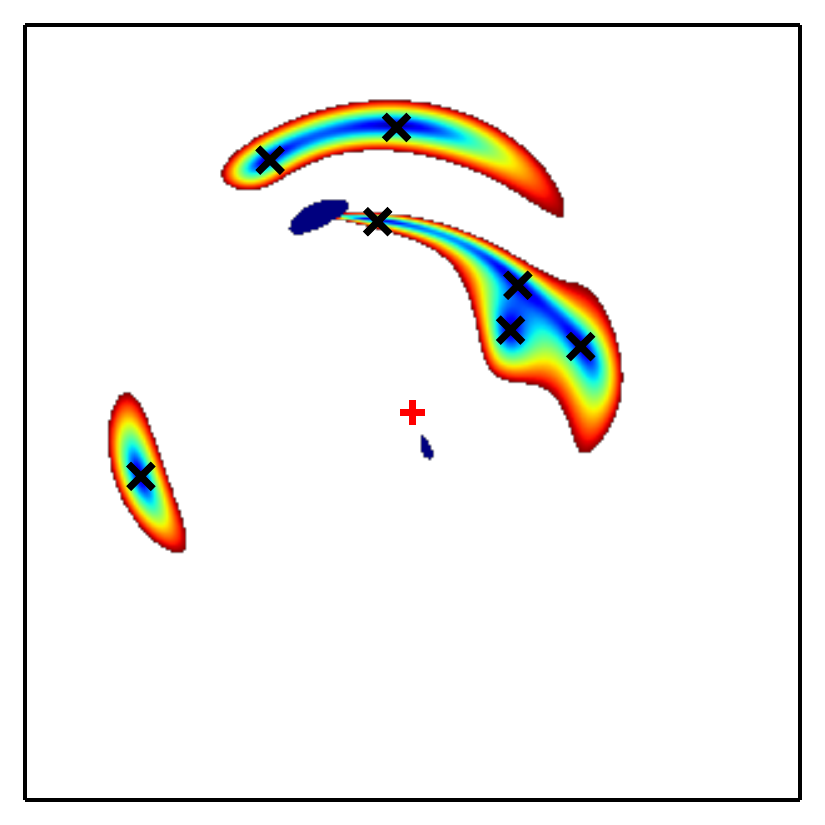}
    \caption {A montage of high multiplicity compound lenses generated from the {\sc LensPop} population of plausible deflectors. The black crosses show the location that images of a point source would form, whilst the central region of an extended source is shown in light blue, with the outer regions shown in red. If the second lens is associated with an extended light profile (assumed to be circular), it would appear as the dark blue region. The optical centre of the first lens is shown as a red cross}
    \label{fig:images_montage}
\end{figure*}


\section*{Acknowledgements}

We are grateful to Charlotte Mason for calculating the size of the compound lens population discoverable by Euclid.

We are grateful to Charlotte Mason and Bob Nichol for useful discussions and comments on the draft manuscript. We are grateful to Matt Auger for the lensing deflections code. We are grateful to Thomas Tram for improvements to our caustic tracing code.

This work has used the LSST sky simulations; we are grateful to Andy Connolly for sharing these, and to Scott Daniel and Gary Burton for enabling access to them.

This project made use of the {\sc LensPop} code which is freely available from \url{github.com/tcollett/LensPop}.

Numerical computations were done on the Sciama High Performance Compute (HPC) cluster which is supported by the ICG, SEPNet and the University of Portsmouth.




\input{references.tex}

\label{lastpage}
\bsp

\end{document}

%% file: macros.tex
\newcommand{\vect}[1]{\boldsymbol{#1}}

\newcommand{\vecarr}[1]{\begin{pmatrix} #1 \end{pmatrix}}

\newcommand{\etal}{et~al.~}
\def\spose#1{\hbox  to 0pt{#1\hss}}  
\newcommand{\lta}{\mathrel{\spose{\lower 3pt\hbox{$\sim$}}\raise  2.0pt\hbox{$<$}}}
\newcommand{\gta}{\mathrel{\spose{\lower  3pt\hbox{$\sim$}}\raise 2.0pt\hbox{$>$}}}

\newcommand{\be}{\begin{equation}}
\newcommand{\ee}{\end{equation}}

\newcommand{\citepeg}[1]{\citep[e.g.][]{#1}}

\newcommand{\kms}{\ifmmode  \,\mathrm {km}\,s^{-1} \else $\,\mathrm{ km\,s^{-1} } $ \fi }
\newcommand{\kpc}{\ifmmode  {\mathrm{kpc}}  \else ${\mathrm{  kpc}}$ \fi  }  
\newcommand{\pc}{\ifmmode  {\mathrm{ pc}}  \else ${\mathrm{ pc}}$ \fi  }  
\newcommand{\Msun}{\ifmmode {\mathrm{ M_{\odot}}} \else ${\mathrm{ M_{\odot}}}$ \fi} 
\newcommand{\Zsun}{\ifmmode {\mathrm{{Z_{\odot}}}} \else ${\mathrm{ Z_{\odot}}}$ \fi} 
\newcommand{\yr}{\ifmmode yr^{-1} \else $yr^{-1}$ \fi} 
\newcommand{\hMsun}{\ifmmode h^{-1}\,\rm M_{\odot} \else $h^{-1}\,mathrm{\ M_{\odot}}$ \fi}

\renewcommand{\etal}{et~al.~}
\def\spose#1{\hbox  to 0pt{#1\hss}}  
\renewcommand{\lta}{\mathrel{\spose{\lower 3pt\hbox{$\sim$}}\raise  2.0pt\hbox{$<$}}}
\renewcommand{\gta}{\mathrel{\spose{\lower  3pt\hbox{$\sim$}}\raise 2.0pt\hbox{$>$}}}

\renewcommand{\be}{\begin{equation}}
\renewcommand{\ee}{\end{equation}}

\newcommand{\bea}{\begin{eqnarray}}
\newcommand{\eea}{\end{eqnarray}}

\newcommand{\comment}[1]{}
\newcommand{\comments}[1]{}

\newcommand{\tom}[1]{#1}

\newcommand{\apj}{ApJ}

\newcommand{\nat}{Nature}

\newcommand{\url}[1]{#1}

%% file: addresses.tex
\def\icg{Institute of Cosmology and Gravitation, University of Portsmouth, Burnaby Rd, Portsmouth, PO1 3FX, UK}

\def\collettemail{\tt thomas.collett@port.ac.uk}